\documentclass[aps,pra,twocolumn,showpacs,floatfix]{revtex4-1}

\usepackage{graphicx}
\usepackage{natbib}
\usepackage{amsmath}
\usepackage{color}
\usepackage[colorinlistoftodos]{todonotes}
\usepackage{mathtools}
\usepackage{mathrsfs}
\usepackage{mathrsfs}

\begin{document}

\title{Lie algebraic approach to quadratic Hamiltonians and the bi-dimensional
charged particle in time-dependent electromagnetic field 
}

 \author{
  V. G. Ibarra-Sierra$^1$,
  J. C. Sandoval-Santana$^1$,
  J.L. Cardoso$^2$
  and
  A. Kunold$^2$
 }
 \affiliation{ 
    \\
   $^1$ Departamento de F\'isica, Universidad Aut\'onoma Metropolitana
   Iztapalapa, Av. San Rafael Atlixco 186, Col. Vicentina,
   09340 M\'exico D.F., M\'exico
   \\
  $^2$ \'Area de F\'isica Te\'orica y Materia Condensada,
   Universidad Aut\'onoma Metropolitana  Azcapotzalco,
   Av. San Pablo 180, Col. Reynosa-Tamaulipas, Azcapotzalco,
   02200 M\'exico D.F., M\'exico 
   } 

\begin{abstract}
We discuss the one-dimensional, general quadratic Hamiltonian and the
bi-dimensional charged particle in time-dependent electromagnetic fields
through the Lie algebraic approach.
Such method consists in finding a set of generators that form a closed
Lie algebra in terms of which it is possible to express the Hamiltonian
and the therefore the evolution operator.
The evolution operator is then the starting point to obtain the propagator
as well as the explicit form of the Heisenberg picture position and momentum operators.
First, the set of generators forming a closed Lie algebra is identified
for the general quadratic Hamiltonian. This algebra is later
extended to study the the Hamiltonian of a charged particle in electromagnetic fields, given the similarities between the terms of these two
Hamiltonians.
\end{abstract}

\maketitle
\section{introduction}
The simple quantum oscillator is the building block of a very large number of well established physical models.
Some of its most widespread applications
are the atomic and molecular bonds that, under certain approximations, can be modelled by quadratic potentials.
The time-dependent general harmonic oscillator (GHO),
the most general version of a simple quantum harmonic
oscillator, is at the heart of many interesting applications as
radio-frequency ion traps.
It consists of a simple harmonic oscillator with
time-varying coefficients, time-dependent linear terms on the
position and momentum operator and
an extra term proportional to the symmetrized product of the position an momentum operators.
It can be described by the quadratic Hamiltonian
\begin{multline}
\hat H=\frac{1}{2}a\left(t\right)\hat p^2+\frac{1}{2}b\left(t\right) \left(\hat x\hat p+\hat p\hat x\right)
+\frac{1}{2}c\left(t\right)\hat x^2\\
+d\left(t\right)\hat p+e\left(t\right)\hat x+g\left(t\right),
\end{multline}
where $\hat x$ and $\hat p$ are the position and momentum operators obeying
the usual commutation relation $\left[\hat x, \hat p\right]=i\hbar$ and
$a$, $b$, $c$, $d$, $e$ and $g$ are in general functions of time.
Since in many cases it possesses exact solutions,
it has turned into a key element to
understanding and modelling a wide variety of physical systems
where potentials are time-dependent.
Specifically the GHO has been applied in diverse branches of physics as
quantum optics\cite{1464-4266-4-3-379,Singh20104685,Mandal2004308},
transport theory in two dimensional electron systems \cite{Inarrea201410,Kunold201378},
quantum field theory \cite{Vergel20091360},
Ions traps (Paul traps) \cite{PhysRevLett.66.527},
laser cooling of trapped ions \cite{PhysRevA.49.421,RevModPhys.75.281,PhysRevA.89.032502},
quantum dissipation (Kanai-Caldirola Hamiltonians)
\cite{caldirola:393,kanai:440,PhysRev.38.815,Um200263,
0305-4470-19-15-024,PhysRevA.68.052108,IbarraSierra201386}, 
and even cosmology \cite{PhysRevD.49.788,Pedrosa2007384}.
One of the main advantages of modelling quantum physical systems
with the GHO is that in many occasions it is exactly
solvable \cite{LopesdeLima20082253}.
The GHO has been studied by diverse mathematical methods such as
the group-theoretical approach \cite{PhysRevA.44.2057},
the path integral approach \cite{PhysRevLett.18.636.2},
unitary transformations \cite{PhysRevLett.66.527,PhysRevA.58.1765},
and  the Lewis and Riesenfeld \cite{lewis:1458}
invariant theory
\cite{PhysRevA.55.3219,PhysRevA.20.550,
0305-4470-34-37-321,LopesdeLima20082253,
PhysRevA.55.4023,PhysRevA.68.052108}.

Besides the GHO, the
time-dependent linear potential (LP),
a particular case of the GHO,
has also received considerable attention also due to the many applications
in fields such as quantum optics, solid state physics,
quantum field theory, molecular physics and quantum chemistry
among others.
It has been established, at least since the 50's, that
the LP's quantum propagator-and also the GHO's propagator-
possess a structure similar to 
the well known propagator for the simple quantum oscillator plus an 
interaction-dependent correction due to the forcing term in the Hamiltonian.
\cite{Merzbarcher3th,Schwinger1951}.
Whereas early studies of the LP relay on
proposed Gaussian-like wave function \cite{Husimi1953}
and standard
variables changes \cite{Popov1970},
recently, the quantum forced harmonic oscillator
has been treated through more powerful methods as
the Lewis and Riesenfeld \cite{lewis:1458} invariant theory
\cite{PhysRevA.63.034102,PhysRevA.68.016101,maamache:1063},
Feynman's path integrals
\cite{kiyoto:777,Khandekar1978, Feynman1948, Feynman1950, Merzbarcher3th},
the generalization of the well known 
ladder operators \cite{Kim1996},
Laplace transform techniques \cite{douglas:199}
and time-space transformation methods \cite{chao:981}.

Similarly to the GHO and the LP, the Hamiltonian describing a particle in
time-dependent electromagnetic fields (CP) has countless applications in many physics fields
such as quantum optics \cite{PhysRevA.46.5885},
single electron quantum dots \cite{PhysRevA.81.052331} and
magneto-transport theory \cite{Inarrea201410,Kunold201378}.
This system has been studied through different methods
that include the Lewis and Riesenfeld \cite{lewis:1458} invariant theory
\cite{PhysRevA.66.024103,PhysRevA.73.016101},
path integral method \cite{0305-4470-17-4-022},
unitary transformation approach \cite{1402-4896-73-6-024,IbarraSierra201386},
and through quadratic invaritants \cite{contentaipjournaljmp52810.10631.3615516}.

The aim of this paper is to apply the Lie algebraic
approach \cite{CPA:CPA3160070404,jmathphys.1.1703993,
PhysRevA.18.89,0305-4470-21-22-015,PhysRevA.87.022116} to
compute the evolution operator of the GHO.
Drawing on these results we also calculate the evolution operator
for the CP Hamiltonian.
Additionally we obtain the propagator and the
explicit form of the Heisenberg picture position and momentum operators.
 
The mass-varying oscillator's evolution operator
was calculated by means of the $SU(2)$
generators in Ref. \cite{0305-4470-21-22-015,Ng1997144}.  The Lie
algebraic approach was also used to study the linear
potential in Ref. \cite{QUA:QUA22781}
and the Kanai-Caldirola Hamiltonian in Ref. \cite{1751-8121-40-5-013}.
However, even though the Lie algebraic approach has been widely used to
treat similar Hamiltonians, it has not been applied  to solve 
specifically neither
the CP nor the GHO Hamiltonians to the extent of our
knowledge.

The paper is organized as follows.
In Sec. \ref{genmethod} we give an overview of the Lie algebraic approach.
First, the time-dependent linear potential serves as an example to sketch
the method and to work out some of the operators that form the Lie algebra in
Sec. \ref{linear potential}.
Second, in Sec. \ref{gen}, we deal with the evolution operator of the most general form of
the quadratic Hamiltonian expanding  the  linear potential  Lie algebra.
With these general results we derive analytical expressions for a
radio frequency ion trap in Sec. \ref{iontrap} and a Kanai-Caldirola forced harmonic oscillator
in Sec. \ref{forced}.
To complete our discussion we extend the Lie algebra of the GHO
by introducing the angular momentum and extra generators. 
Finally
we treat the Hamiltonian of a 2D charged particle
in time-dependent electromagnetic fields exploiting the similarities of its Hamiltonian
with GHO one.
With the general expressions hereby obtained we
compute analytical expressions for a charged particle
in time-varying magnetic field in Sec. \ref{bt} and
time-dependent electric fields in Sec. \ref{et}.
We give the final conclusions in Sec. \ref{conclusions}.
The Lie algebra generators, their commutation relations and
their structure constants are presented in Appendix \ref{liealgebra}.
Their corresponding unitary transformations are
presented in Appendix \ref{unitarytransformations} along with their transformation rules
and propagators.

\section{Overview to the Lie algebraic approach}\label{genmethod}

A Hamiltonian is said to have a dynamical algebra if it can be
expressed as the linear combination
\begin{equation}
\hat H=\sum_{i=1}^{n} a_i\left(t\right)\hat \lambda_i,
\end{equation}
where $a_i\left(t\right)$ are real functions of time and the
set of Hermitian operators
$\Lambda=\left\{\hat \lambda_1, \hat \lambda_2,\dots \hat \lambda_n\right\}$
forms a closed Lie algebra $\mathcal L$. $\mathcal L$ is characterized by
the structure
constant $c_{i,j,k}$ in the commutor
\begin{equation}
\left[\hat \lambda_i,\hat \lambda_j\right]=i\hbar\sum_{k=1}^n c_{i,j,k}\hat \lambda_k .\label{strucons}
\end{equation}
In the sections to follow we show that the LP, the GHO and the CP
Hamiltonians have dynamical algebras by identifying their generators
and the corresponding structure constants.

The Lie algebraic approach \cite{CPA:CPA3160070404,jmathphys.1.1703993,
PhysRevA.18.89,0305-4470-21-22-015,PhysRevA.87.022116}  relays on the fact
that the evolution operators of such Hamiltonians
can be expressed in either of the following forms
\begin{eqnarray}
\hat {\mathcal U}\left(t\right) &=& \exp\left[i\sum_{i=1}^n \alpha_i\left(t\right)\hat \lambda_i\right],\\
\hat {\mathcal U}\left(t\right) &=& \prod_{i=1}^n \exp\left[ i\beta_i\left(t\right)\hat 
\lambda_i\right],\label{unit0}
\end{eqnarray}
where the transformation parameters
$\alpha_i$ and $\beta_i$ are differentiable functions of time yet to be determined.

We first consider
Schr\"odinger's equation
\begin{equation}\label{ec.schrodinger}
\hat{H} \left\vert \psi \left( t\right) \right\rangle =\hat{p}_t \left\vert\
\psi\left(t\right)\right\rangle,
\end{equation}
where $\hat{p}_t =i\hbar \partial /\partial t$,
and, conveniently, we introduce the Floquet operator \cite{Heinzpeter2006}
\begin{equation}\label{floquet}
\hat{\mathcal H} =\hat{H} -\hat{p}_t ,
\end{equation}
that allows to write Schr\"odinger equation in the rather compact form  
\begin{equation}\label{shro}
\hat{\mathcal H} \left\vert \psi \left(t \right) \right\rangle= 0.
\end{equation}
Using Eq. (\ref{unit0})
let us now assume that there is a set of unitary transformations
\begin{equation}\label{met:eq4}
\mathcal G=\left\{ \hat{ U}_{1}, 
\hat{ U}_{2},\dots \hat{ U}_{n}\right\},
\end{equation}
with time-dependent
transformation parameters
$\beta_1\left(t\right)$, $\beta_2\left(t\right)$, $\dots$, $\beta_n\left(t\right)$
in the form of Eq. (\ref{unit0})
such that the application of
\begin{equation}
\hat{U}=\hat{ U}_{n} \dots \hat{ U}_{2}\hat{ U}_{1},
\end{equation}
to the Floquet operator reduces it
to the energy operator removing the Hamiltonian part
as shown below
\begin{equation}
\hat{U}\hat{\mathcal H}\hat{U}^{\dagger}= -\hat{p}_t.
\label{met:eq3}
\end{equation}
We further assume that the
explicit forms of the transformation rules of
$\hat U_i \hat x\hat U_i^\dagger$, $\hat U_i \hat p\hat U_i^\dagger$
and $\hat U_i \hat p_t\hat U_i^\dagger$ for any
unitary transformation in $\mathcal G$  are known. The explicit form
of the transformation rules of the unitary transformations used in this paper
are presented in
Appendix \ref{unitarytransformations}.
Conditions on the transformation parameters must be
found so as to satisfy Eq. (\ref{met:eq3}).
As it is shown in Section \ref{gen}, two different sets of unitary operators
corresponding to the same Hamiltonian might comply
with Eq. (\ref{met:eq3}) meaning that there may be two or more different
ways of arriving to the same evolution operator.

If such a transformation does exist, the 
Schr\"odinger equation takes the form
\begin{equation}
\hat{U} \hat{\mathcal H} \hat{U}^{\dagger} \hat{U} \left\vert \psi \left( 
t\right)\right\rangle = -\hat{p}_t \left(\hat{U} \left\vert \psi \left( 
t\right)\right\rangle\right)=0.\label{met:eq0}
\end{equation}
Reminding that $\hat p_t$ is $\hbar$ times a time derivative, it is easy to
see that $\hat{U}\left\vert\psi\left(t\right)\right\rangle$ is a constant ket,
i. e.
\begin{equation}
\hat{U} \left\vert \psi \left( t\right) \right\rangle = \left\vert \psi \left(
0\right) \right\rangle . \label{met:eq1}
\end{equation}
Therefore the evolution of a 
quantum state $\psi$ can be easily calculated by multiplying the previous 
equation by the inverse of $U$ ($U^{-1}=U^\dag$) getting
\begin{equation}
\left\vert \psi \left( t\right) \right\rangle = \hat{U}^{\dagger} \left\vert 
\psi \left( 0\right) \right\rangle.\label{met:eq2}
\end{equation}
This equation states that obtained unitary operator $\hat{U}^{\dagger}$
is in fact the time evolution operator
i. e. $\hat{U}^{\dagger}=\hat {\mathcal U}$.

The Green function, or the propagator, is calculated as usual in terms of 
the evolution operator as
\begin {multline}
G \left( x,t; x^\prime,0 \right) =\left\langle x\left\vert \hat U^{\dagger} 
\right\vert x^\prime\right\rangle
=\left\langle x\left\vert \hat U_1^{\dagger}\hat U_2^{\dagger}\dots
\hat U_n^{\dagger} 
\right\vert x^\prime\right\rangle\\
=\int dx_1\int dx_2\dots \int dx_{n-1}
\left\langle x\left\vert \hat U_1^{\dagger}\right\vert x_1\right\rangle\\
\times\left\langle x_1\left\vert\hat U_2^{\dagger}\right\vert x_2\right\rangle
\dots
\left\langle x_{n-1}\left\vert\hat U_n^{\dagger} 
\right\vert x^\prime\right\rangle.\label{prop}
\end{multline}
The explicit form of the position and momentum operators in the
Heisenberg picture
may be worked out from the transformation rules
as
\begin{eqnarray}
x_H\left(t\right) &=& \hat U \hat x \hat U^{\dag},\label{heisx}\\
p_H\left(t\right) &=& \hat U \hat p \hat U^{\dag}.\label{heisp}
\end{eqnarray}

%
%
\section{Linear potential}\label{linear potential}
To illustrate the use of the Lie algebraic approach,  we analyze the
solution of the one dimensional Schr\"odinger equation
of a particle with variable mass subject to a time-dependent linear potential
\cite{PhysRevA.63.034102,PhysRevA.68.016101,
maamache:1063,QUA:QUA22781,QUA:QUA22781}.
The time-dependent mass term allows us to study dissipation
in Kanai-Caldirola-like Hamiltonians \cite{caldirola:393,kanai:440,PhysRev.38.815}.
Such Hamiltonian is given by
\begin{equation}
H=\frac{1}{2m\left(t\right)}{\hat p}^2-f\left(t\right){\hat x}\label{ham:lin},
\end{equation}
where the mass $m\left(t\right)$ and the force $f\left(t\right)$ depend arbitrarily on time.
From now on we drop their time-dependence except in special cases.
The difficulty in finding the evolution operator for the time-dependent
potential becomes evident when one computes the commutor of the Hamiltonian
at two different moments in times $t_1$ and $t_2$
\begin{equation}
\left[\hat H\left(t_1\right),\hat H\left(t_2\right)\right]=
i\hbar \hat p\left[\frac{f\left(t_2\right)}{m\left(t_1\right)}
-\frac{f\left(t_1\right)}{m\left(t_2\right)}\right].
\end{equation}
In general, the last commutor does not vanish therefore
the Hamiltonian (\ref{ham:lin})
does not allow the evolution operator to be written in the simple form
$\exp\left[-i\int_0^t\hat H\left(t\right)dt\right]$ requiring a
different approach.

Now we turn our attention to the generators of the
Hamiltonian (\ref{ham:lin}).
At first glance, the set of operators $\left\{\hat x, \hat p^2\right\}$
seems like the
right choice for $\mathcal L$, however, a closer look at the
commutation relations reveals that in order to close the algebra
we must also include $\hat 1$ and $\hat p$.
Thereby the whole set is given by
$\hat \lambda_1=\hat 1$, $\hat \lambda_2=\hat x$,
$\hat \lambda_3=\hat p$, and $\hat \lambda_4=\hat p^2$,
where $\hat 1$ is the identity operator.
In appendix \ref{gens:lin} we present the commutors
and structure constants
of these generators; it is shown that in fact
the algebra, exhibited by $\hat \lambda_1$-$\hat \lambda_4$, is closed.

Even though this set of operators in principle
guarantees that the evolution operator
should be given by
$\hat U=\exp\left(\beta_1\right)\exp\left(\beta_2 \hat x\right)
\exp\left(\beta_3\hat p\right)\exp\left(\beta_4\hat p^2\right)$, we proceed
applying each generator's unitary transformation  stepwisely. 
Our first goal is to eliminate the linear term on the position operator
$\hat x$ therefore
we first apply the translation in space and momentum
(see Appendix \ref{unitarytranslation})
generated by
$\hat \lambda_1=\hat 1$, $\hat \lambda_2=\hat x$ and $\hat \lambda_3=\hat p$
given by
\begin{equation}
{\hat U}_1\left(t\right)=\exp\left[\frac{i}{\hbar}S\left(t\right)\right]
\exp\left[\frac{i}{\hbar}\Pi\left(t\right)\hat x\right]
\exp\left[\frac{i}{\hbar}\lambda\left(t\right)\hat p\right],\label{u1:lin}
\end{equation}
where $S\left(t\right)$, $\Pi\left(t\right)$ and $\lambda\left(t\right)$
are the real and differentiable time-dependent transformation parameters
$\beta_1$, $\beta_2$ and $\beta_3$.
The transformation rules for (\ref{u1:lin}) are given by
\begin{eqnarray}
{\hat U}_1{\hat p}_t{\hat U}_1^{\dagger} &=&
{\hat p}_t+\dot S-\dot\lambda\Pi+\dot\Pi\hat x+\dot\lambda\hat p ,\\
{\hat U}_1\hat x{\hat U}_1^\dagger &=& \hat x+\lambda,\\
{\hat U}_1\hat p{\hat U}_1^\dagger &=& \hat p-\Pi,
\end{eqnarray}
where an overdot denotes a time derivative.
Under this transformation,
the Floquet operator is transformed into
\begin{multline}
{\hat U}_1\left(H-{\hat p}_t\right){\hat U}_1^{\dagger}=
\frac{1}{2m}\left({\hat p}-\Pi\right)^2-f\left(t\right)\left({\hat x}+\lambda\right)\\
-\left({\hat p}_t+\dot S-\dot\lambda\Pi+\dot\Pi\hat x+\dot\lambda \hat p\right)\\
=\frac{1}{2m}{\hat p}^2-{\hat p}_t-\left(\frac{\Pi}{m}+\dot\lambda\right)\hat p
-\left(f+\dot\Pi\right)\hat x\\
-\left(\dot S+f\lambda-\dot\lambda \Pi-\frac{\Pi^2}{2m}\right).\label{tra:lin:u1}
\end{multline}
In order to vanish the linear terms in $\hat x$ and $\hat p$
we must set
the following conditions on the transformation parameters
\begin{eqnarray}
\frac{\Pi}{m}+\dot\lambda &=& 0,\label{u1:lin:eq1}\\
f+\dot\Pi &=& 0, \label{u1:lin:eq2}
\end{eqnarray}
with initial conditions $\lambda\left(0\right)=\Pi\left(0\right)=0$ in order for
${\hat U}_1$ to be equal to the identity operator at $t=0$ i. e. 
${\hat U}_1\left(0\right)= \hat 1$.

Equally, to cancel the independent terms, we must set
\begin{equation}
\dot S+f\lambda-\dot\lambda \Pi-\frac{\Pi^2}{2m}=0,\label{u1:lin:eq3}
\end{equation}
with initial condition $S\left(0\right)=0$.
Immediately we notice the parallel between
Eqs. (\ref{u1:lin:eq1}) and (\ref{u1:lin:eq2})
and the Hamilton equations
of motion for the classical analog of  (\ref{ham:lin}).
Moreover, if we collect the independent terms in Eq. (\ref{tra:lin:u1})
and define the classical Lagrangian
\begin{equation}
 L\left(t\right)\equiv\frac{\Pi^2}{2m}+\dot\lambda \Pi-f\lambda,
\end{equation}
its Euler equations yield the conditions imposed on
the transformation parameters (\ref{u1:lin:eq1}) and (\ref{u1:lin:eq2})
\begin{eqnarray}
\frac{d}{dt}\frac{\partial  L}{\partial \dot\lambda}-\frac{\partial  L}{\partial \lambda} &=&
\dot\Pi +f=0,\\
\frac{d}{dt}\frac{\partial  L}{\partial\dot\Pi}-\frac{\partial  L}{\partial\Pi} &=& 
-\frac{\Pi}{m}-\dot\lambda=0.
\end{eqnarray}
The analogy goes even further when we notice that
Eq. (\ref{u1:lin:eq3}) is in fact the standard definition
of the classical action
\begin{equation}
S=\int_0^tds L\left(s\right).
\end{equation}
Once these conditions are set, the original Floquet operator is simplified into
the one of a free particle 
\begin{equation}
{\hat U}_1\left(H-{\hat p}_t\right){\hat U}_1^{\dagger}=\frac{1}{2m}{\hat p}^2-{\hat p}_t.
\end{equation}
As it is desirable that all the terms from the Hamiltonian are eliminated,
it is clear that the last transformation should be the one generated
by $\hat \lambda_4=\hat p^2$ (see Appendix \ref{unitaryp2})
\begin{equation}
{\hat U}_2\left(t\right)=\exp\left[\frac{i}{2\hbar}\beta\left(t\right){\hat p}^2\right],
\end{equation}
that yields the  transformation rules
\begin{eqnarray}
{\hat U}_2{\hat p}_t{\hat U}_2^{\dagger} &=& {\hat p}_t+\frac{\dot\beta}{2}{\hat p}^2,\\
{\hat U}_2{\hat x}{\hat U}_2^{\dagger} &=& \hat x+\beta \hat p,\\
{\hat U}_2{\hat p}{\hat U}_2^{\dagger} &=& \hat p.
\end{eqnarray}
Application of this transformation to the Floquet operator gives
\begin{equation}
{\hat U}_2
{\hat U}_1\left(H-{\hat p}_t\right)
{\hat U}_1^{\dagger}{\hat U}_2^{\dagger}
=\frac{1}{2}\left(\frac{1}{m}-\dot\beta\right){\hat p}^2-{\hat p}_t.
\end{equation}
By establishing the restriction
\begin{equation}
\frac{1}{m}-\dot\beta=0,\label{u1:lin:eq4}
\end{equation}
with the initial condition $\beta\left(0\right)=0$ in order to make ${\hat U}_2\left(0\right)=\hat 1$,
the Floquet operator is finally reduced to the energy operator
\begin{equation}
{\hat U}_2
{\hat U}_1\left(H-{\hat p}_t\right)
{\hat U}_1^{\dagger}{\hat U}_2^{\dagger}
=-{\hat p}_t.
\end{equation}
Therefore, by Eqs. (\ref{met:eq1}) and (\ref{met:eq2})
the evolution operator is given by
\begin{multline}
\hat{ U}^{\dag}\left(t\right)={\hat U}_1^\dag\left(t\right){\hat U}_2^\dag\left(t\right)
=
\exp\left[-\frac{i}{\hbar}S\left(t\right)\right]\\
\times
\exp\left[-\frac{i}{\hbar}\lambda\left(t\right)\hat p\right]
\exp\left[-\frac{i}{\hbar}\Pi\left(t\right)\hat x\right]\\
\times
\exp\left[-\frac{i}{2\hbar}\beta\left(t\right){\hat p}^2\right].
\end{multline}
Solving the differential equations (\ref{u1:lin:eq1})-(\ref{u1:lin:eq3})
and (\ref{u1:lin:eq4}) one obtains the transformation parameters
\begin{eqnarray}
\lambda\left(t\right) &=& 
\int_0^t \frac{ds}{m\left(s\right)} \int_0^s
dr f\left(r\right),\label{lin:par1}\\
\Pi\left(t\right) &=& -\int_0^t ds f\left(s\right),\\
\beta\left(t\right) &=& \int_{0}^{t} \frac{ds}{m\left(s\right)}\label{lin:par2}.
\end{eqnarray}

Using Eqs. (\ref{heisx}) and (\ref{heisp}) we compute
the position and momentum operator in the Heisenberg picture
\begin{eqnarray}
{\hat x}_H &=& \hat x +\beta\left(t\right) \hat p +\lambda\left(t\right),\\
{\hat p}_H &=& \hat p -\Pi\left(t\right).
\end{eqnarray} 
Hence, for given force $f\left(t\right)$ and mass $m\left(t\right)$ functions,
one can easily determine
all the transformation parameters through Eqs. (\ref{lin:par1})-(\ref{lin:par2})
and plug this solutions into the propagator and Heisenberg picture
space and momentum operators.

Finally, from Eq. (\ref{prop})
and the propagators shown in Appendices \ref{unitarytranslation}
and \ref{unitaryp2},
the propagator for the LP can be expressed as
\begin{multline}
G\left(x,t;x^\prime,0\right)=\int dx_1\left\langle x\left\vert \hat U_1^\dag\right\vert x_1\right\rangle
\left\langle x_1\left\vert \hat U_2^\dag\right\vert x^\prime\right\rangle\\
=\frac{1}{\sqrt{2\pi\hbar \beta\left(t\right)}}
\exp\left[-\frac{i}{\hbar}S\left(t\right)\right]
\exp\left\{-i\frac{\Pi\left(t\right)}{\hbar}\left[x-\lambda\left(t\right)\right]\right\}\\
\exp\left\{\frac{i}{2\hbar\beta\left(t\right)}\left[x-x^\prime-\lambda\left(t\right)\right]^2\right\}.
\end{multline}

%
%
\section{General quadratic Hamiltonian}\label{gen}
With the earlier treatment we can handle the
GHO Hamiltonian with time-dependent coefficients.
We follow two different procedures to obtain the unitary operator
for the GHO Hamiltonian in order to study two different
special cases:
a radio frequency ion trap and the Kanai-Caldirola Hamiltonian
of a forced harmonic oscillator.

We start with the most general Hamiltonian
\begin{multline}
H=\frac{1}{2}a\left(t\right)\hat p^2+\frac{1}{2}b\left(t\right)
\left(\hat x\hat p+\hat p\hat x\right)
+\frac{1}{2}c\left(t\right)\hat x^2\\
+d\left(t\right)\hat p+e\left(t\right)\hat x+g\left(t\right),
\label{ham:quadratic}
\end{multline}
with arbitrary time-dependent coefficients $a\left(t\right)$, $b\left(t\right)$,
$c\left(t\right)$, $d\left(t\right)$, $e\left(t\right)$ and $g\left(t\right)$.
It is quite clear
from the structure of (\ref{ham:quadratic})
that the closed algebra corresponding to this Hamiltonian should be given
by the set of operators
$\hat\lambda_1=\hat 1$, $\hat\lambda_2=\hat x$, $\hat\lambda_3=\hat p$,
$\hat\lambda_4=\hat x^2$, $\hat\lambda_5=\hat p^2$,
$\hat\lambda_6=\hat x\hat p+\hat p\hat x$.
In Appendix \ref{gens:quadratic} we present the commutation
relations and the structure constants for these generators.
 
We first aim to remove the independent terms and the ones
proportional to $\hat x$ and $\hat p$. 
We thus apply the translation in space and momentum
shown in Appendix \ref{unitarytranslation}
generated
by $\hat\lambda_1=\hat 1$, $\hat\lambda_2=\hat x$ and $\hat \lambda_3=\hat p$
given by
\begin{equation}
{\hat U}_1\left(t\right)=\exp\left[\frac{i}{\hbar}S\left(t\right)\right]
\exp\left[\frac{i}{\hbar}\Pi\left(t\right)\hat x\right]
\exp\left[\frac{i}{\hbar}\lambda\left(t\right)\hat p\right].
\label{gen:u1}
\end{equation}
The transformation rules for (\ref{gen:u1}) are given by
\begin{eqnarray}
{\hat U}_1{\hat p}_t{\hat U}_1^{\dagger} &=&
   {\hat p}_t+\dot S-\dot\lambda\Pi+\dot\Pi\hat x+\dot\lambda\hat p ,\label{transrule1}\\
{\hat U}_1\hat x{\hat U}_1^\dagger &=& \hat x+\lambda,\\
{\hat U}_1\hat p{\hat U}_1^\dagger &=& \hat p-\Pi \label{transrule3}.
\end{eqnarray}
Applying (\ref{transrule1})-(\ref{transrule3}) to (\ref{ham:quadratic})
The transformed Floquet operator takes the form
\begin{multline}
{\hat U}_1\left(\hat H-{\hat p}_t\right){\hat U}_1^\dagger 
=\frac{1}{2}a{\hat p}^2+\frac{1}{2}b\left(\hat x\hat p+\hat p\hat x\right)
+\frac{1}{2}c{\hat x}^2-{\hat p}_t\\
+\hat x\left(c\lambda -b\Pi+e-\dot\Pi\right)\\
+\hat p\left(-a\Pi+b\lambda+d-\dot\lambda\right)\\
+g-\dot S+\frac{1}{2}a\Pi^2+\frac{1}{2}c\lambda^2-b\lambda\Pi-d\Pi+e\lambda+\dot\lambda\Pi.
\end{multline}
It is possible vanish
the independent and linear terms in $\hat x$ and $\hat p$
by imposing the following restriction on the
transformation parameters
\begin{eqnarray}
 c\lambda -b\Pi+e-\dot\Pi &=& 0,\label{u1:eq1} \\
 -a\Pi+b\lambda+d-\dot\lambda &=& 0,\label{u1:eq2}
 \end{eqnarray}
 \begin{equation}
\dot S = g+\frac{1}{2}a\Pi^2+\frac{1}{2}c\lambda^2-b\lambda\Pi-d\Pi
                 +e\lambda+\dot\lambda\Pi, \label{u1:eq3}
\end{equation}
with initial conditions
$S\left(0\right)=\lambda\left(0\right)=\Pi\left(0\right)=0$
in order to guarantee that $U_1\left(0\right)$ is the identity
operator at $t=0$.

Once again, we observe that Eqs. (\ref{u1:eq1})-(\ref{u1:eq2}) are the classical
Euler equations corresponding to the Lagrangian
\begin{equation}
 L=\frac{1}{2}a\Pi^2+\frac{1}{2}c\lambda^2-b\lambda\Pi-d\Pi+e\lambda+\dot\lambda\Pi+g.
\end{equation}
We can readily obtain Eqs. (\ref{u1:eq1}) and (\ref{u1:eq2}) from the Euler equations
\begin{eqnarray}
\frac{d}{dt}\frac{\partial L}{\partial \dot \lambda}
   -\frac{\partial L}{\partial \lambda}
    &=&  \dot\Pi-c\lambda +b\Pi-e=0,\\
\frac{d}{dt}\frac{\partial L}{\partial \dot \Pi}
   -\frac{\partial  L}{\partial \Pi}
    &=& \dot\lambda+a\Pi-b\lambda-d=0,
\end{eqnarray}
and Eq. (\ref{u1:eq3}) yields the very well known relation for the action
\begin{equation}
S=\int_0^t ds  L\left(s\right).
\end{equation}

Imposing conditions (\ref{u1:eq1})-(\ref{u1:eq3}),
the transformed Floquet operator reduces to the quadratic form
\begin{multline}
{\hat U}_1\left(\hat H-{\hat p}_t\right){\hat U}_1^\dagger \\
=\frac{1}{2}a{\hat p}^2+\frac{1}{2}b\left(\hat x\hat p+\hat p\hat x\right)
+\frac{1}{2}c{\hat x}^2-{\hat p}_t .\label{gen:u1:eq1}
\end{multline}

As pointed out earlier, there may be different sets of unitary
transformations that reduce the Floquet operator to $\hat p_t$.
To illustrate two possible solutions for the
evolution operator of the general quadratic Hamiltonian, at this
point, we take two different calculation paths. The first one is shorter
but requires the solution of Riccati differential equation
whereas the second path
is more involved but in a wide variety of physical situations avoids
solving Riccati differential equation through the use of Arnold transformation.

Now we consider
the dilation generated by $\hat \lambda_6=\hat x\hat p+\hat p\hat x$
(see Appendix \ref{unitarydilation}) given by
\begin{equation}
{\hat U}_2\left(t\right)=\exp\left[\frac{i}{2\hbar}\gamma\left(t\right)
\left({\hat x}{\hat p}+{\hat p}{\hat x}\right)\right],\label{gen:u2}
\end{equation}
that yields the following transformation rules
\begin{eqnarray}
{\hat U}_2{\hat p}_t{\hat U}_2^{\dagger} &=&
   {\hat p}_t+\frac{1}{2}\dot\gamma
\left({\hat x}{\hat p}+{\hat p}{\hat x}\right),\label{u2:tr1}\\
{\hat U}_2\hat x{\hat U}_2^\dagger &=&{\mathrm e}^{\gamma} \hat x,\label{u2:tr2}\\
{\hat U}_2\hat p{\hat U}_2^\dagger &=& {\mathrm e}^{-\gamma} \hat p \label{u2:tr3}.
\end{eqnarray}
The application of the dilation yields the following transformed Floquet operator
\begin{multline}
{\hat U}_2{\hat U}_1\left(\hat H-{\hat p}_t\right){\hat U}_1^\dagger{\hat U}_2^{\dagger}
=\frac{1}{2}a{\mathrm e}^{-2\gamma} {\hat p}^2
+\frac{1}{2}c{\mathrm e}^{2\gamma} {\hat x}^2\\
+\frac{1}{2}\left(b-\dot\gamma\right)\left(\hat x\hat p+\hat p\hat x\right)-{\hat p}_t.
\label{u2:tr6}
\end{multline}
Although doing $\dot \gamma=b$ to remove the term proportional to
$\left(\hat x\hat p+\hat p\hat x\right)$ would seem to simplify
the Floquet operator, it is more convenient to leave $\gamma$ as a
free parameter that will be useful later on.

\subsection{First path}\label{firstpath}
We take the calculation from Eq. (\ref{u2:tr6}).
In this path it is convenient to set the $\gamma$ parameter by doing
\begin{equation}
\mathrm{e}^{2\gamma}=a\Delta,\label{gamma:path1}
\end{equation}
and $\Delta=1/a\left(0\right)=1/a_0$ in order to make the
dilation $\hat U_2$ equal to the identity operator at
$t=0$.
This seemingly arbitrary definition of $\gamma$ and $\Delta$ will
prove to be a key step in simplifying Riccati equation into a linear
second order differential equation.

It is possible to get a notable simplification by applying
the unitary transformation generated by $\hat \lambda_4=\hat x^2$
(see Appendix \ref{unitaryx2})
given by
\begin{equation}
U_3=\exp\left[\frac{i}{2\hbar}\alpha\left(t\right)\Delta \hat x^2\right],
\label{firstpath:u1}
\end{equation}
that yields the following transformation rules
\begin{eqnarray}
\hat U_3 \hat p_t \hat U^{\dag}_3 &=& \hat p_t
+\dot \alpha\frac{\Delta}{2}\hat x^2,\\
\hat U_3 \hat x \hat U^{\dag}_3 &=& \hat x,\\
\hat U_3 \hat p \hat U^{\dag}_3 &=& \hat p-\alpha \Delta \hat x.
\end{eqnarray}
After carrying the transformations above the Floquet operator takes the form
\begin{multline}
\hat U_3\hat U_2{\hat U}_1\left(\hat H-{\hat p}_t\right)
{\hat U}_1^{\dagger}\hat U_2^{\dagger}\hat U_3^{\dagger} \\
=
\frac{1}{2}
\left[
a\mathrm{e}^{-2\gamma}\Delta^2 \alpha^2
-2\left(b-\dot\gamma\right)\Delta \alpha
+c\mathrm{e}^{2\gamma}
-\Delta\dot\alpha
\right]
{\hat x}^2\\
+\frac{1}{2}a\mathrm{e}^{-2\gamma}\ {\hat p}^2
+\frac{1}{2}\left(b-\dot \gamma-a\mathrm{e}^{-2\gamma}\Delta\alpha\right)
\left(\hat x\hat p+\hat p\hat x\right)
-{\hat p}_t .
\end{multline}
It is desirable that the term proportional to $\hat x^2$ vanish,
hence we restrict the values of the $\alpha$ parameter
by setting the condition
\begin{equation}
\Delta\dot\alpha=
a\mathrm{e}^{-2\gamma}\Delta^2 \alpha^2
-2\left(b-\dot\gamma\right)\Delta \alpha
+c\mathrm{e}^{2\gamma}.\label{fp:eq1}
\end{equation}
This is a Riccati differential equation of the form
\begin{equation}
y'\left(x\right)=q_0\left(x\right)
+q_1\left(x\right)y\left(x\right)+q_2\left(x\right)y^2\left(x\right),
\label{ricatti}
\end{equation}
with $\alpha=y$, $q_0=c\mathrm{e}^{2\gamma}/\Delta$, $q_1=-2\left(b-\dot\gamma\right)$ and
$q_2=a\mathrm{e}^{-2\gamma}\Delta$.
Note that applying the restriction (\ref{gamma:path1})
and doing the variable change $\alpha=-\dot u/u$,
the Riccati equation is turned
into the simpler linear second order differential equation
\begin{equation}
\ddot u+\left(2b-\frac{\dot a}{a}\right)\dot u+ca u=0.
\label{linearsndord}
\end{equation}  
If Eq. (\ref{fp:eq1})- or equivalently Eq. (\ref{linearsndord})- hold the transformed Floquet operator
becomes
\begin{multline}
\hat U_3\hat U_2{\hat U}_1\left(\hat H-{\hat p}_t\right)
{\hat U}_1^{\dagger}\hat U_2^{\dagger}\hat U_3^{\dagger} 
=\frac{1}{2}a\mathrm{e}^{-2\gamma}\ {\hat p}^2\\
+\frac{1}{2}\left(b-\dot \gamma-a\mathrm{e}^{-2\gamma}\Delta\alpha\right)
\left(\hat x\hat p+\hat p\hat x\right)
-{\hat p}_t .
\end{multline}
The dilation generated by $\hat \lambda_6=\hat x\hat p+\hat p\hat x$
\begin{equation}
\hat U_4\left(t\right)=\exp\left[\frac{i}{2\hbar}\phi\left(t\right)
\left({\hat x}{\hat p}+{\hat p}{\hat x}\right)\right],
\label{firstpath:u4}
\end{equation}
seems the right choice for the next transformation
since it trivially commutes
with its generator $\hat x\hat p+\hat p\hat x$.
This transformation produces the transformation
rules given by
\begin{eqnarray}
{\hat U}_4{\hat p}_t{\hat U}_4^{\dagger} &=&
   {\hat p}_t+\frac{1}{2}\dot\phi
\left({\hat x}{\hat p}+{\hat p}{\hat x}\right),\\
{\hat U}_4\hat x{\hat U}_4^\dagger &=&{\mathrm e}^{\phi} \hat x,\\
{\hat U}_4\hat p{\hat U}_4^\dagger &=& {\mathrm e}^{-\phi} \hat p .
\end{eqnarray}
Application of the dilation yields the transformed
Floquet operator
\begin{multline}
\hat U_4
\hat U_3\hat U_2{\hat U}_1\left(\hat H-{\hat p}_t\right)
{\hat U}_1^{\dagger}\hat U_2^{\dagger}\hat U_3^{\dagger} \hat U_4^{\dagger}
=\frac{1}{2}a\mathrm{e}^{-2\left(\gamma+\phi\right)}\ {\hat p}^2\\
+\frac{1}{2}\left(b-\dot \gamma-\dot\phi-a\mathrm{e}^{-2\gamma}\Delta\alpha\right)
\left(\hat x\hat p+\hat p\hat x\right)
-{\hat p}_t  .
\end{multline}
In order to eliminate the term proportional to ${\hat x}{\hat p}+{\hat p}{\hat x}$
we set 
\begin{equation}
\dot\phi=b-\dot \gamma-a\mathrm{e}^{-2\gamma}\Delta\alpha,\label{first:phi:u4}
\end{equation}
obtaining the Floquet operator for a free particle with variable mass
\begin{multline}
\hat U_4
\hat U_3\hat U_2{\hat U}_1\left(\hat H-{\hat p}_t\right)
{\hat U}_1^{\dagger}\hat U_2^{\dagger}\hat U_3^{\dagger} \hat U_4^{\dagger}\\
=\frac{1}{2}a\mathrm{e}^{-2\left(\gamma+\phi\right)}\ {\hat p}^2
-{\hat p}_t  .
\end{multline}

It is clear that in order to remove the remaining term we must
apply the transformation generated by $\hat \lambda_5=\hat p^2$
that can be expressed as follows
\begin{equation}
\hat U_5\left(t\right)=\exp\left[
\frac{i}{2\hbar }\beta\left(t\right)\frac{\hat p^2}{\Delta}
\right],\label{firstpath:u5}
\end{equation}
with the transformation rules given by
\begin{eqnarray}
\hat U_5\hat p_t\hat U_5^{\dagger} &=& \hat p_t+\dot\beta\frac{1}{2\Delta}\hat p^2,\\
\hat U_5\hat x\hat U_5^{\dagger} &=& \hat x+\beta\frac{\hat p}{\Delta},\\
\hat U_5\hat p\hat U_5^{\dagger} &=& \hat p.
\end{eqnarray}
Therefore,
under this transformation the Floquet operator takes the form
\begin{multline}
\hat U_5
\hat U_4\hat U_3\hat U_2{\hat U}_1
\left(\hat H-{\hat p}_t\right){\hat U}_1^{\dagger}\hat U_2^{\dagger}
\hat U_3^{\dagger}\hat U_4^{\dagger}
\hat U_5^{\dagger} \\
=\frac{1}{2}\left[a\mathrm{e}^{-2\left(\gamma+\phi\right)}-\frac{\dot\beta}{\Delta}\right]{\hat p}^2
-{\hat p}_t  .
\end{multline}
Imposing the following restriction on the transformation parameter
\begin{equation}
\dot\beta=\Delta a\mathrm{e}^{-2\left(\gamma+\phi\right)}, \label{first:beta:u5}
\end{equation}
the Floquet operator is finally rendered
into the energy operator i. e.
\begin{equation}
\hat U_5
\hat U_4\hat U_3\hat U_2{\hat U}_1\left(\hat H-{\hat p}_t\right)
{\hat U}_1^{\dagger}\hat U_2^{\dagger}
\hat U_3^{\dagger}\hat U_4^{\dagger}
\hat U_5^{\dagger} \\
=-{\hat p}_t .
\end{equation}
According to Eqs. (\ref{met:eq0}) and (\ref{met:eq1}) this product of transformations
is precisely the evolution operator
\begin{multline}
{\hat U}^{\dagger}={\hat U}_1^{\dagger}\hat U_2^{\dagger}
\hat U_3^{\dagger}\hat U_4^{\dagger}\hat U_5^{\dagger}=
\exp\left(-\frac{i}{\hbar}\lambda\hat p\right)
\exp\left(-\frac{i}{\hbar}\Pi\hat x\right)\\
\times\exp\left(-\frac{i}{\hbar}S\right)
\exp\left[-\frac{i}{2\hbar}\gamma
\left({\hat x}{\hat p}+{\hat p}{\hat x}\right)\right]\\
\exp\left(-\frac{i}{2\hbar}\alpha\Delta \hat x^2\right)
\exp\left[-\frac{i}{2\hbar}\phi
\left({\hat x}{\hat p}+{\hat p}{\hat x}\right)\right]\\
\times\exp\left(-\frac{i}{2\hbar }\beta\frac{\hat p^2}{\Delta}\right).
\label{first:evolution}
\end{multline}
Collecting the results above,
the classical equations of motion for the position and momentum
are obtained from the Euler equations (\ref{u1:eq1}) and (\ref{u1:eq2})
\begin{eqnarray}
\dot\Pi &=&  c\lambda -b\Pi+e \label{pi:dot}, \\
\dot\lambda &=&  b\lambda-a\Pi+d \label{lambda:dot},
\end{eqnarray}
and the corresponding classical action can be calculated by substituting
the explicit forms of $\lambda$ and $\Pi$
into (\ref{u1:eq3}) and integrating
\begin{multline}
S = \int_0^t ds\left[g\left(s \right)+\frac{1}{2}a\left(s \right)\Pi^2\left(s \right)
+\frac{1}{2}c\left(s \right)\lambda\left(s \right)^2
\right. \\ 
-b\left(s \right)\lambda\left(s\right)\Pi\left(s \right)
-d\left(s \right)\Pi\left(s \right)
\\ \left.
 +e\left(s \right)\lambda\left(s \right)+\dot\lambda\left(s \right)\Pi\left(s \right)\right].
\end{multline}
To complete the remaining parameters, from (\ref{gamma:path1})
we set $\gamma=\ln(a/a_0)/2$ in order to simplify the
Riccati Eq. (\ref{fp:eq1}). Next, the $\alpha$ parameter may be
integrated either from (\ref{fp:eq1}) or (\ref{linearsndord}).
Then, the $\phi$ and $\beta$ parameters are calculated by direct
integration of the ordinary differential equations
(\ref{first:phi:u4}) and (\ref{first:beta:u5})
\begin{eqnarray}
\phi\left(t\right) &=&-\gamma\left(t\right)\nonumber\\
&&+ \int_0^{t}ds\left[b\left(s\right)
-a\left(s\right)\mathrm{e}^{-2\gamma\left(s\right)} \Delta \alpha\left(s\right)\right],
\label{phi:int}\\
\beta\left(t\right) &=&  \int_0^{t}ds 
\Delta a\left(s\right)\mathrm{e}^{-2\left[\gamma\left(s\right)+ \phi \left(s\right)\right]}.
\label{beta:int}
\end{eqnarray}

Putting the explicit form of the evolution operator (\ref{first:evolution})
into Eqs. (\ref{heisx}), (\ref{heisp})
the Heisenberg picture position and momentum operators can be expressed as
\begin{equation}
\left[\begin{array}{l}
\hat x_{H}\left(t\right) \\
\hat p_{H}\left(t\right) 
\end{array}\right]=
\mathbf{M}
\left[\begin{array}{l}
\hat x \\
\hat p 
\end{array}\right]+
\left[\begin{array}{r}
\lambda \\
-\Pi 
\end{array}\right],\label{heis:ptdm}
\end{equation}
where
\begin{equation}
\mathbf{M}=\left[\begin{array}{ll}
G_{qq} & G_{qp} \\
G_{pq} & G_{pp} 
\end{array}\right],
\end{equation}
and
\begin{eqnarray}
G_{qq} &=& \mathrm{e}^{\phi+\gamma},\label{coe:heis1}\\
G_{qp} &=& \frac{\beta}{\Delta}\mathrm{e}^{\phi+\gamma},\\
G_{pq} &=&  -\alpha \Delta \mathrm{e}^{\phi-\gamma},\\
G_{pp} &=&
\mathrm{e}^{-\phi-\gamma}-\alpha\beta\mathrm{e}^{\phi-\gamma}.
\label{coe:heis4}
\end{eqnarray}
The matrix $\mathbf{M}$ satisfies the symplectic conditions
inherited from the unitary transformations
$\mathbf{M}^\top i \sigma_y \mathbf{M}=\mathbf{M} i \sigma_y \mathbf{M}^\top=i \sigma_y$ with $\sigma_y$ the Pauli matrix.
Additionally it complies
with $\det\mathbf{M}=1$. The three previous conditions ensure that
the commutation relations between position and momentum operators
are preserved during the system's evolution,
namely $\left[\hat x_H(t),\hat p_H(t)\right]=i\hbar$.

Using Eq. (\ref{prop})
and the Green functions of the five unitary transformations
in Appendix \ref{unitarytransformations} we readily integrate
the propagator
\begin{multline}
G\left(x,t;x^{\prime},0\right)=\sqrt{\frac{\Delta}{2\pi\hbar \beta}}
\exp\left[-i\frac{S\left(t\right)}{\hbar}\right]
\exp\left(-\frac{\phi+\gamma}{2}\right)\\
\times
\exp\left[i\frac{\Delta \mathrm{e}^{-2\gamma}}{2\hbar}
\left(\frac{ \mathrm{e}^{-2\phi}}{\beta}-\alpha\right)
\left(x-\lambda\right)^2
\right]
\exp\left[i\frac{\Delta}{2\hbar \beta}\left(x^{\prime}\right)^2\right]\\
\times
\exp\left[-
i\left(\frac{\Delta \mathrm{e}^{-\phi-\gamma} }{\hbar \beta}x^{\prime}+\frac{\Pi}{\hbar}\right)
\left(x-\lambda\right)
\right].\label{firstpath:prop}
\end{multline}

\subsection{Second path}\label{secondpath}
Here we follow an alternative path to the one in
the previous section.
In this path we start the calculation
from Eq. (\ref{u2:tr6}) but instead of
(\ref{gamma:path1})
we impose the following restriction on the $\gamma$ parameter
\begin{equation}
\mathrm{e}^{2\gamma}=\Delta\sqrt{\frac{a}{c}}=
\sqrt{\frac{c\left(0\right)a}{a\left(0\right)c}}=
\sqrt{\frac{c_0a}{a_0c}},\label{gamma:rest}
\end{equation}
where $\Delta=\sqrt{c_0/a_0}$ in order to make the
dilation $\hat U_2$ equal to the identity operator at
$t=0$.

First we apply Arnold's transformation
\begin{equation}
{\hat U}_3\left(t\right)=\exp\left[\frac{i}{2\hbar}\phi\left(t\right)
\left(\Delta {\hat x}^2+\frac{1}{\Delta}{\hat p}^2\right)
\right],\label{gen:u3}
\end{equation}
where $\phi$ is the transformation parameter.
The transformation rules for (\ref{gen:u3}) are
\begin{eqnarray}
{\hat U}_3{\hat p}_t{\hat U}_3^{\dagger} &=&
   {\hat p}_t+\frac{1}{2}\dot\phi\left(\frac{1}{\Delta}{\hat p}^2
  + \Delta {\hat x}^2\right),\\
{\hat U}_3\hat x{\hat U}_3^\dagger &=& \hat x \cos\phi+\frac{1}{\Delta}\hat p\sin\phi,\\
{\hat U}_3\hat p{\hat U}_3^\dagger &=& \hat p \cos\phi-\Delta \hat x \sin \phi.
\end{eqnarray}
Note that Arnold's transformation is generated by a
linear combination of $\hat \lambda_4=\hat x^2$ and $\hat \lambda_5=\hat p^2$.
Under this transformation the Floquet operator takes the form
\begin{multline}
{\hat U}_3{\hat U}_2{\hat U}_1\left(\hat H-{\hat p}_t\right)
{\hat U}_1^\dagger{\hat U}_2^{\dagger} {\hat U}_3^{\dagger}=\\
=\frac{1}{2\Delta}\left[
\sqrt{ac}-\dot \phi+\left(b-\dot\gamma\right)\sin 2\phi
 \right]{\hat p}^2\\
+\frac{\Delta}{2}\left[
\sqrt{ac}-\dot \phi-\left(b-\dot\gamma\right)\sin 2\phi
\right]{\hat x}^2\\
+\frac{1}{2}\left(b-\dot\gamma\right)\cos 2\phi\left(\hat x\hat p+\hat p\hat x\right) -{\hat p}_t.
\end{multline}

By restricting Arnold's transformation parameter by the relation
\begin{equation}
\dot\phi=\sqrt{ac},\label{u3:eq5}
\end{equation}
the Floquet operator reduces to the following quadratic form
proportional to $b-\dot\gamma$
\begin{multline}
{\hat U}_3{\hat U}_2{\hat U}_1\left(\hat H-{\hat p}_t\right)
{\hat U}_1^\dagger{\hat U}_2^{\dagger} {\hat U}_3^{\dagger}\\
=\frac{1}{2}\left(b-\dot\gamma\right)
\left[\left(\frac{\hat p^2}{\Delta}-\Delta\hat x^2\right)\sin{2\phi}
\right.\\
+ \left(\hat x\hat p+\hat p\hat x\right)\cos2\phi
\bigg]-\hat p_t.\label{gen:u30}
\end{multline}
Certain cases where 
\begin{equation}
b-\dot\gamma=0,\label{simple}
\end{equation}
specially those where $b=\dot\gamma=0$, lead to physically meaningful
systems such as a variable mass charged particle in constant magnetic field.
It is thus worthwhile to treat them separately.
If condition (\ref{simple}) is fulfilled
the Floquet operator is completely reduced to the energy operator $\hat p_t$
implying that the evolution operator is simply given by
\begin{equation}
\hat U^{\dagger}={\hat U}_1^{\dagger}{\hat U}_2^{\dagger}{\hat U}_3^{\dagger}.
\end{equation}
However, in order to consider cases where $b-\dot\gamma \ne 0 $ we must move
on to the next transformation. We consider the unitary transformation generated
by $\hat \lambda_4=\hat x^2$ given by
\begin{equation}
\hat U_4\left(t\right)=\exp\left[\frac{i}{2\hbar}\alpha\left(t\right)\Delta \hat x^2\right],
\label{gen:u4}
\end{equation}
with the following transformation rules
\begin{eqnarray}
\hat U_4 \hat p_t \hat U^{\dag}_4 &=& \hat p_t+\dot \alpha\frac{\Delta}{2}\hat x^2,\\
\hat U_4 \hat x \hat U^{\dag}_4 &=& \hat x,\\
\hat U_4 \hat p \hat U^{\dag}_4 &=& \hat p-\alpha \Delta \hat x.
\end{eqnarray}
The application of this transformation to the Floquet operator in Eq. (\ref{gen:u30})
yields 
\begin{multline}
{\hat U}_4{\hat U}_3{\hat U}_2{\hat U}_1\left(\hat H-{\hat p}_t\right)
{\hat U}_1^\dagger{\hat U}_2^{\dagger} {\hat U}_3^{\dagger} {\hat U}_4^{\dagger}\\
=\frac{1}{2}\left(b-\dot\gamma\right)\left[\frac{\hat p^2}{\Delta}\sin 2\phi
+ \left(\hat x\hat p+\hat p\hat x\right)\left(\cos2\phi-\alpha\sin2\phi\right)\right]\\
+\frac{1}{2}\left\{\left(b-\dot\gamma\right)\left[\left(\alpha^2-1\right)\sin2\phi
-2\alpha\cos 2\phi\right]
-\dot \alpha\right\}\Delta \hat x^2.\label{gen:u40}
\end{multline}
To eliminate the terms proportional to $\hat x^2$ we set
the following restriction on $\alpha$
\begin{equation}
\dot \alpha=\left(b-\dot\gamma\right)\left[\left(\alpha^2-1\right)\sin2\phi
-2\alpha\cos 2\phi\right].\label{u4:eq5}
\end{equation}
This is newly a Riccati differential equation of the form (\ref{ricatti})
with $q_0=-\left(b-\dot\gamma\right)\sin 2\phi$, $q_1=2\left(b-\dot\gamma\right)\cos 2\phi$
and $q_2=\left(b-\dot\gamma\right)\sin 2\phi$ .

After the condition (\ref{u4:eq5}) has been set,
the Floquet operator (\ref{gen:u40}) takes the form
\begin{multline}
{\hat U}_4{\hat U}_3{\hat U}_2{\hat U}_1\left(\hat H-{\hat p}_t\right)
{\hat U}_1^\dagger{\hat U}_2^{\dagger} {\hat U}_3^{\dagger} {\hat U}_4^{\dagger}\\
=\frac{1}{2}\left(b-\dot\gamma\right)\left[\frac{\hat p^2}{\Delta}\sin 2\phi
+ \left(\hat x\hat p+\hat p\hat x\right)\left(\cos2\phi-\alpha\sin2\phi\right)\right]\\
-\hat p_t .
\end{multline}

It is convenient to set the nearly last transformation to be a dilation
of the form
\begin{equation}
{\hat U}_5\left(t\right)=
\exp\left[\frac{i}{2\hbar}\varphi\left(t\right)\left(\hat x\hat p+\hat p\hat x\right) \right],
\label{gen:u5}
\end{equation}
since it just multiplies the $\hat p^2$ term by a factor $\exp\left(-2\varphi\right)$
and yields an additional $\dot\varphi \left(\hat x\hat p+\hat p\hat x\right)$ term that allows to
cancel the $\left(b-\dot\gamma\right)\left(\hat x\hat p+\hat p\hat x\right)\left(\cos2\phi-\alpha\sin2\phi\right)$.
Indeed, transcribing the transformation rules from (\ref{u2:tr1})-(\ref{u2:tr3})
\begin{eqnarray}
{\hat U}_5{\hat p}_t{\hat U}_5^{\dagger} &=&
   {\hat p}_t+\frac{1}{2}\dot\varphi
\left({\hat x}{\hat p}+{\hat p}{\hat x}\right),\label{u5:tr1}\\
{\hat U}_5\hat x{\hat U}_5^\dagger &=&{\mathrm e}^{\varphi} \hat x,\label{u5:tr2}\\
{\hat U}_5\hat p{\hat U}_5^\dagger &=& {\mathrm e}^{-\varphi} \hat p, \label{u5:tr3}
\end{eqnarray}
and applying $\hat U_5$ to the previous Floquet operator we obtain
the above described terms
\begin{multline}
{\hat U}_5{\hat U}_4{\hat U}_3{\hat U}_2{\hat U}_1
\left(\hat H-{\hat p}_t\right)
{\hat U}_1^\dagger{\hat U}_2^{\dagger} {\hat U}_3^{\dagger}{\hat U}_4^{\dagger}
{\hat U}_5^{\dagger}\\
=\frac{1}{2\Delta}\left(b-\dot\gamma\right){\rm e}^{-2\varphi}\hat p^2\sin 2\phi\\
+\frac{1}{2}
 \left(\hat x\hat p+\hat p\hat x\right)\left[\left(b-\dot\gamma\right)
\left(\cos2\phi-\alpha\sin2\phi\right)-\dot\varphi\right]\\-\hat p_t .
\end{multline}
By imposing
\begin{equation}
\dot\varphi=\left(b-\dot\gamma\right)\left(\cos2\phi-\alpha\sin2\phi\right),\label{u5:tr7}
\end{equation}
the Floquet operator is reduced to the one of
a free particle with variable mass
\begin{multline}
{\hat U}_5{\hat U}_4{\hat U}_3{\hat U}_2{\hat U}_1
\left(\hat H-{\hat p}_t\right)
{\hat U}_1^\dagger{\hat U}_2^{\dagger} {\hat U}_3^{\dagger}{\hat U}_4^{\dagger}
{\hat U}_5^{\dagger}\\
=\frac{1}{2\Delta}\left(b-\dot\gamma\right){\rm e}^{-2\varphi}\hat p^2\sin 2\phi
-\hat p_t .
\end{multline}

Evidently, the last transformation to be used is
the one generated by $\hat \lambda_5=\hat p^2$
\begin{equation}
\hat U_6\left(t\right)=\exp\left[\frac{i}{2\hbar }\beta\left(t\right)\frac{\hat p^2}{\Delta}\right],
\label{gen:u6}
\end{equation}
with transformation rules
\begin{eqnarray}
\hat U_6\hat p_t\hat U_6^{\dagger} &=& \hat p_t+\dot\beta\frac{1}{2\Delta}\hat p^2,\\
\hat U_6\hat x\hat U_6^{\dagger} &=& \hat x+\beta\frac{\hat p}{\Delta},\\
\hat U_6\hat p\hat U_6^{\dagger} &=& \hat p.
\end{eqnarray}
In this case, the Floquet operator takes the form
\begin{multline}
{\hat U}_6
{\hat U}_5{\hat U}_4{\hat U}_3{\hat U}_2{\hat U}_1
\left(\hat H-{\hat p}_t\right)
{\hat U}_1^\dagger{\hat U}_2^{\dagger} {\hat U}_3^{\dagger}{\hat U}_4^{\dagger}
{\hat U}_5^{\dagger}{\hat U}_6^{\dagger}\\
=\frac{1}{2}\left[
\left(b-\dot\gamma\right){\rm e}^{-2\varphi} \sin 2\phi
-\dot\beta
\right]\frac{\hat p^2}{\Delta}
-\hat p_t .
\end{multline}
The Floquet operator above
can be reduced to the energy operator
$\hat p_t$ by imposing the following condition on $\beta$
\begin{equation}
\dot \beta=\left(b-\dot\gamma\right){\rm e}^{-2\varphi} \sin 2\phi.\label{u6:tr5}
\end{equation}
We have thus arrived to the form (\ref{met:eq3}) and therefore
by collecting
(\ref{gen:u1}), (\ref{gen:u2}), (\ref{gen:u3}), (\ref{gen:u4}), (\ref{gen:u5})
and (\ref{gen:u6}) the evolution operator is given by
\begin{multline}
\hat U^{\dagger}={\hat U}_1^\dagger{\hat U}_2^{\dagger} {\hat U}_3^{\dagger}{\hat U}_4^{\dagger}
{\hat U}_5^{\dagger}{\hat U}_6^{\dagger}\\
=\exp\left(-\frac{i}{\hbar}S\right)
\exp\left(-\frac{i}{\hbar}\lambda\hat p\right)
\exp\left(-\frac{i}{\hbar}\Pi\hat x\right)\\
\times
\exp\left[-\frac{i}{2\hbar}\gamma
\left({\hat x}{\hat p}+{\hat p}{\hat x}\right)\right]\\
\times
\exp\left[-\frac{i}{2\hbar}\phi
\left(\Delta {\hat x}^2+\frac{1}{\Delta}{\hat p}^2\right)\right]
\exp\left(-\frac{i}{2\hbar}\alpha\Delta \hat x^2\right)\\
\times
\exp\left[-\frac{i}{2\hbar}\varphi\left(\hat x\hat p+\hat p\hat x\right) \right]
\exp\left(-\frac{i}{2\hbar}\beta\frac{\hat p^2}{\Delta}\right)\label{gen:u}.
\end{multline}
The transformation parameters must be calculated from the
restrictions (\ref{u1:eq1}), (\ref{u1:eq2}),  (\ref{u1:eq3}),  (\ref{gamma:rest}),
 (\ref{u3:eq5}), (\ref{u4:eq5}), (\ref{u5:tr7}) and (\ref{u6:tr5}).

By successively  applying the six transformations above to $\hat x$ and $\hat p$
we can workout the Heisenberg picture position and momentum operators
newly obtaining the symplectic form
\begin{equation}
\left[\begin{array}{l}
\hat x_{H}\left(t\right) \\
\hat p_{H}\left(t\right) 
\end{array}\right]=
\mathbf{M}
\left[\begin{array}{l}
\hat x \\
\hat p 
\end{array}\right]+
\left[\begin{array}{r}
\lambda \\
-\Pi 
\end{array}\right],
\end{equation}
where
\begin{equation}
\mathbf{M}=\left[\begin{array}{ll}
G_{qq} & G_{qp} \\
G_{pq} & G_{pp} 
\end{array}\right],
\end{equation}
but instead, in this case the matrix elements are given by
\begin{multline}
G_{qq}\left(t\right)=\left(\cos\phi-\alpha \sin\phi\right){\mathrm e}^{\gamma+\varphi} \\
=\left(\cos\phi-\alpha \sin\phi\right)\sqrt[4]{\frac{c_0 a}{a_0 c}}{\mathrm e}^{\varphi},
\label{gqq}
\end{multline}
\begin{multline}
G_{qp}\left(t\right)=
\left[\left(\beta\cos\phi-\alpha\beta\sin\phi\right){\mathrm e}^{\varphi}+\sin\phi{\mathrm e}^{-\varphi} \right]
\frac{{\mathrm e}^{\gamma}}{\Delta}\\
=\sqrt[4]{\frac{a_0 a}{c_0 c}}
\left[\left(\beta\cos\phi-\alpha\beta\sin\phi\right){\mathrm e}^{\varphi}+\sin\phi{\mathrm e}^{-\varphi} \right],
\label{gqp}
\end{multline}
\begin{multline}
G_{pq}\left(t\right)=-\left(\alpha\cos\phi+\sin\phi\right)\Delta{\mathrm e}^{\varphi-\gamma}\\
=-\left(\alpha\cos\phi+\sin\phi\right)
\sqrt[4]{\frac{c_0 c}{a_0 a}}{\mathrm e}^{\varphi}.\label{gpq}
\end{multline}
\begin{multline}
G_{pp}\left(t\right)=-
\left[\left(\beta\sin\phi+\alpha\beta\cos\phi\right){\mathrm e}^{\varphi}-\cos\phi{\mathrm e}^{-\varphi} 
\right]
{\mathrm e}^{-\gamma}\\
=-\left[\left(\beta\sin\phi+\alpha\beta\cos\phi\right){\mathrm e}^{\varphi}-\cos\phi{\mathrm e}^{-\varphi} 
\right]
\sqrt[4]{\frac{a_0c}{c_0a}}.\label{gpp}
\end{multline}
Even though the structure of the matrix $\mathbf{M}$ is radically different
from the one obtained in the first path
[see Eqs. (\ref{coe:heis1})-(\ref{coe:heis4})], it also
satisfies the symplectic conditions. 
From Eq. (\ref{prop}) and the propagators presented in
Appendix  \ref{unitarytransformations} we can obtain
the propagator associated to (\ref{gen:u}) as
\begin{multline}
G\left(x,t;x^\prime,0\right)=\int dx_1dx_2dx_3dx_4 dx_5\\
\times
\left\langle x\left\vert \hat U^{\dagger}_1\right\vert x_1\right\rangle
\left\langle x_1\left\vert \hat U^{\dagger}_2\right\vert x_2\right\rangle
\left\langle x_2\left\vert \hat U^{\dagger}_3\right\vert x_3\right\rangle\\
\times
\dots
\left\langle x_5\left\vert \hat U^{\dagger}_6\right\vert x_6\right\rangle\\
=\sqrt{\frac{\Delta^2}{4i\pi\hbar^2\beta l \sin\delta}}
\exp\left(-i\frac{S}{\hbar}-\frac{\varphi+\gamma}{2}\right)
\exp\left[
iw \left(x-\lambda\right)^2
\right.\\
\left.
+iu\left(x^\prime\right)^2
+i\left(qx^\prime-\frac{\Pi}{\hbar}\right)\left(x-\lambda\right)
\right],
\end{multline}
where, for the sake of brevity, we have defined the following functions
\begin{eqnarray}
u &=& \frac{\Delta}{2\hbar \beta}\left(1+\frac{\Delta \mathrm{e}^{-2\varphi}}{2\hbar \beta l}\right),
\label{w:function}\\
w &=& \frac{\Delta\mathrm{e}^{-2\varphi}}{2\hbar\sin\phi}\left(\cos\phi+\frac{\Delta}{2\hbar 
l\sin\phi}\right),\\
q &=& \frac{\Delta^2\mathrm{e}^{-\left(\varphi+\gamma\right)}}{2\hbar^2 l\beta\sin\phi},\\
l &=& \Delta\frac{\beta \left(\alpha-\cot\phi\right)- \mathrm{e}^{-2\varphi}}{2\hbar \beta}
\label{l:function}.
\end{eqnarray}

\subsection{Radio frequency ion trap}\label{iontrap}
Here we assume that the trap potential
can be decomposed into a static and a time-dependent part
that varies sinusoidally at the drive radio-frequency
$\omega$  \cite{RevModPhys.75.281}.
Thus, the ion Hamiltonian is given by
\begin{multline}
\hat H=\frac{\hat p_x^2}{2m}+\frac{\hat p_y^2}{2m}+\frac{\hat p_z^2}{2m}
+\frac{1}{2}\left(K_x +k_x\cos\omega t\right)\hat x^2\\
+\frac{1}{2}\left(K_y +k_y\cos\omega t\right)\hat y^2
+\frac{1}{2}\left(K_z +k_z\cos\omega t\right)\hat z^2.
\end{multline}
The parameters
$k_x$, $k_y$, $k_z$, $K_x$, $K_y$ and $K_z$
are restricted being that the electrical potential has to fulfil Laplace equation. A possible
choice is $-\left(K_x+K_y\right)=K_z>0$ and $k_x=-k_y$.
The separability of the previous Hamiltonian allows us
to work the $x$, $y$ and $z$ coordinates independently.
Thereby $a_i=1/m$, $b_i=0$, $c_i=K_i+k_i\cos\omega t$, $d_i=e_i=g_i=0$ where
$i=x,y,z$.
The solution to the ordinary differential equations (\ref{pi:dot}) and (\ref{lambda:dot})
with initial conditions $\Pi_i\left(0\right)=\lambda_i\left(0\right)$
is $\Pi_i=\lambda_i=0$ and thus, the action is $S=0$.
When $q_2=1$ Riccati Eq. (\ref{ricatti}) is known to reduce to
a second order linear
equation by making $\alpha_i=-\dot u_i/u_i$ . Therefore,
we set $\Delta a_i \exp\left(-2\gamma_i\right)=1$
by doing $\Delta_i=m$ and $\gamma_i=0$. We are left with the
Mathieu differential equation
\begin{equation}
\ddot u_i+\frac{1}{m}\left(K_i+k_i\cos\omega t\right)u_i=0,
\end{equation}
whose solution is given by
\begin{equation}
u_i=A\ C\left(\frac{4K_i}{m\omega^2},-\frac{2k_i}{m\omega^2},\frac{\omega t}{2}\right),
\end{equation}
where $A$ is a constant and $C\left(a,q,z\right)$ is the Mathieu
cosine function that
complies with $C\left(a,q,0\right)=1$ and its derivative $C^{\prime}\left(a,q,0\right)=0$.
The parameter is therefore given by
\begin{equation}
\alpha_i=
-\frac{\omega C^{\prime}\left(\frac{4K_i}{m\omega^2},-\frac{2k_i}{m\omega^2},\frac{\omega t}{2}\right)}
{2C\left(\frac{4K_i}{m\omega^2},-\frac{2k_i}{m\omega^2},\frac{\omega t}{2}\right)}.
\end{equation}
By substituting the previous result in (\ref{phi:int}) we get
\begin{multline}
\phi_{i}=-\int_0^t ds \alpha_i\left(s\right)=\int_0^t ds \frac{\dot u_i\left(s\right)}{u_i\left(s\right)}\\
=\ln\left[C\left(\frac{4K_i}{m\omega^2},-\frac{2k_i}{m\omega^2},\frac{\omega t}{2}\right)\right].
\end{multline}
The last parameter is
\begin{equation}
\beta_i=T\left(\frac{4K_i}{m\omega^2},-\frac{2k_i}{m\omega^2},\frac{\omega t}{2}\right)=
\int_0^t\frac{ds}{C^2\left(\frac{4K_i}{m\omega^2},-\frac{2k_i}{m\omega^2},\frac{\omega s}{2}\right)}.
\end{equation}
Gathering the results above, the Heisenberg picture operators are given by
\begin{eqnarray}
\hat{x}_{iH}\left(t\right) &=&C_i\left(\frac{\omega t}{2}\right)\hat x_
+\frac{1}{m}C_i\left(\frac{\omega t}{2}\right)T_i\left(\frac{\omega t}{2}\right)
\hat p_i,
\\
\hat{p}_{iH}\left(t\right) &=&
\left[\frac{1}{C_i\left(\frac{\omega t}{2}\right)}
+\frac{\omega }{2}C^{\prime}_i\left(\frac{\omega t}{2}\right)
T_i\left(\frac{\omega t}{2}\right)
\right]
\hat p_i
 \nonumber \\
&&+
\frac{m\omega}{2}C^{\prime}_i\left(\frac{\omega t}{2}\right)
\hat x_i,
\end{eqnarray}
where, for the sake of simplicity we have defined
$C_i\left(\omega t/2\right)=C\left(4K_i/m\omega^2,-2k_i/m\omega^2,\omega t/2\right)$
and $T_i\left(\omega t/2\right)
=\int_0^tdsC^{-2}\left(4K_i/m\omega^2,-2k_i/m\omega^2,\omega s/2\right)$.
Finally, the propagator is given by
\begin{multline}
G\left(x,y,z,t;x^{\prime},y^{\prime},z^{\prime},0\right)=\\
\left(\frac{m}{2\pi\hbar}\right)^{3/2}\prod_{i=x,y,z}
\frac{1}{\sqrt{C_i\left(\frac{\omega t}{2}\right)
T_i\left(\frac{\omega t}{2}\right)}}\\
\times
\exp\left\{i\frac{m}{2\hbar}\sum_{i=x,y,z}
\left[\frac{1}
{C_i^2\left(\frac{\omega t}{2}\right)T_i\left(\frac{\omega t}{2}\right)}
+\frac{\omega C_i^{\prime}\left(\frac{\omega t}{2}\right)}{2C_i\left(\frac{\omega t}{2}\right)}\right]
x_i^2
\right\}\\
\times\exp\left\{
i\frac{m}{2\hbar}\sum_{i=x,y,z}
\frac{1}{T_i\left(\frac{\omega t}{2}\right)}
\left[\left(x_i^{\prime}\right)^2-
\frac{2xx^{\prime}}{C_i\left(\frac{\omega t}{2}\right)}\right]
\right\}.
\end{multline}

\subsection{Forced harmonic oscillator with varying mass}\label{forced}
The harmonic oscillator with varying
mass Hamiltonian is a useful theoretical tool to study quantum dissipation
\cite{caldirola:393,kanai:440}. 
It has been treated by diverse methods including
Feynman integrals \cite{PhysRevA.58.1765},
and the Lie algebraic
approach \cite{0305-4470-21-22-015}. In the latter
the Hamiltonian was expressed by means of the
three $SU(2)$ generators.
In contrast, the forced harmonic oscillator needs
a larger set of generators because of the linear potential
terms.
The
expression for the Kanai-Caldirola Hamiltonian of the forced harmonic
oscillator is 
\begin{multline}
\hat H=\frac{\mathrm{e}^{-t/\tau}}{2m}\hat p^2
+\frac{\mathrm{e}^{t/\tau}}{2}m\omega_0^2 \hat x^2\\
-\mathrm{e}^{t/\tau}\left(F_0+F_1\sin\omega_1t\right)\hat x.
\end{multline}
Therefore, $a=\exp\left(-t/\tau\right)/m$,
$c=\omega_0^2m\exp\left(t/\tau\right)$,
$e=-\exp\left(t/\tau\right)\left(F_0+F_1\sin\omega_1t\right)$ and $b=d=g=0$.
Restricting ourselves to the case of over-damping, i. e. $4\tau^2\omega_0^2<1$,
the $\lambda$ and $\Pi$ parameters can be obtained from the solution
of the ordinary differential Eqs. (\ref{pi:dot}) and (\ref{lambda:dot}) that yield
the standard solutions for the classical damped harmonic oscillator
\begin{multline}
\lambda\left(t\right)=\frac{F_0}{m\omega_0^2}\left(1-\mathrm{e}^{-t/2\tau}\cosh\Omega t
-\frac{\mathrm{e}^{-t/2\tau}}{2\tau\Omega}\sinh\Omega t\right)\\
+\frac{F_1}{m\left[\tau^2\left(\omega_1^2-\omega_0^2\right)^2+\omega_1^2\right]}\\
\times
\left[\tau\omega_1\mathrm{e}^{-t/2\tau}\cosh\Omega t
+\frac{\omega_1}{2\Omega}\left(1+2\tau^2\omega_1^2-2\tau^2\omega_0^2\right)\mathrm{e}^{-t/2\tau}\right.\\
\left.\times\sinh\Omega t -\tau\omega_1\cos\omega_1 t
+\frac{ }{ }\tau^2\left(\omega_0^2-\omega_1^2\right)\sin\omega_1 t\right],
\label{lambda:damp}
\end{multline}
\begin{multline}
\Pi\left(t\right)=- \frac{F_0}{\Omega}\mathrm{e}^{t/2\tau}\sinh\Omega t
+\frac{F_1}{\left[\tau^2\left(\omega_1^2-\omega_0^2\right)^2+\omega_1^2\right]}\\
\times\left[
\tau^2\omega_1\left(\omega_0^2-\omega_1^2\right)\mathrm{e}^{t/2\tau}
\left(\cosh\Omega t -\mathrm{e}^{t/2\tau}\cos\omega_1t\right)\right.\\
\left.+\frac{\tau\omega_1}{2\Omega}\left(\omega_0^2+\omega_1^2\right)
\mathrm{e}^{t/2\tau}\sinh\Omega t
-\tau\omega_1^2\mathrm{e}^{t/\tau}\sin\omega_1 t\right],
\label{pi:damp}
\end{multline}
where $\Omega=\sqrt{\left\vert 1-4\tau^2\omega_0^2\right\vert}/2\tau$.

Now we turn to the $\alpha$, $\phi$ and $\beta$ parameters. As in the
previous example, (\ref{fp:eq1}) is the key equation we have to solve first.
To do so, we set $\gamma=-t/2\tau$, $\Delta=m$ and do the variable change
$\alpha=-\dot u/u$ rendering Riccati equation in the form
of a damped harmonic oscillator
\begin{equation}
\ddot u+\frac{1}{\tau}\dot u+\omega_0^2u=0.
\end{equation}
Under over-damping conditions, the solution for the previous differential equation is
\begin{equation}
\alpha\left(t\right)=\frac{1}{2\tau}\frac{1-4\tau^2\Omega^2}{1+2\tau\Omega \coth\Omega t}.
\label{alpha:damp}
\end{equation}
Integrating (\ref{phi:int}) and (\ref{beta:int}) we obtain the remaining parameters
\begin{eqnarray}
\phi\left(t\right) &=& \ln\left(\cosh\Omega t+\frac{1}{2\tau\Omega}\sinh\Omega t\right),
\label{phi:damp}\\
\beta\left(t\right) &=& \frac{2\tau}{1+2\tau\Omega\coth\Omega t}.
\label{beta:damp}
\end{eqnarray}
The under-damped harmonic oscillator parameters are obtained by doing
 $\Omega\rightarrow i\Omega$
 in (\ref{lambda:damp}), (\ref{pi:damp}), (\ref{alpha:damp}), (\ref{phi:damp})
 and (\ref{beta:damp}).
Note that the three previous results are comparable to the ones
obtained in Ref. \cite{0305-4470-21-22-015} by using the $SU(2)$ generators.
Substituting the explicit forms of the parameters into
(\ref{heis:ptdm}) we obtain the Heisenberg picture position and
momentum operators
\begin{eqnarray}
\hat x_H\left(t\right) &=&
\left(\cosh\Omega t+\frac{1}{2\tau\Omega}\sinh\Omega t\right)\mathrm{e}^{-t/2\tau}\hat x\nonumber \\
&&+\frac{\mathrm{e}^{-t/2\tau}}{m\Omega}\sinh\Omega t\ \hat p+\lambda ,\\
\hat p_H\left(t\right) &=&
\frac{m}{4\tau^2\Omega}\left(4\tau^2\Omega^2-1\right)\mathrm{e}^{t/2\tau}\sinh\Omega t\ \hat x
\nonumber \\
&&+\frac{\mathrm{e}^{t/2\tau}}{2\tau\Omega}\left(2\tau\Omega\cosh\Omega t-\sinh\Omega t\right)\hat p-\Pi .
\end{eqnarray}

Finally, introducing the explicit form of the parameters into (\ref{firstpath:prop})
the propagator can be expressed as
\begin{multline}
G\left(x,t;x^{\prime},0\right)=\sqrt{\frac{m\Omega}{2\pi\hbar\sinh\Omega t}}
\ \mathrm{e}^{-iS/\hbar}\ \mathrm{e}^{t/4\tau}\\
\times\exp\left[
-i\frac{m}{4\hbar \tau}\mathrm{e}^{t/\tau}\left(1-2\tau\Omega\coth\Omega t\right)
\left(x-\lambda\right)^2
\right]\\
\times\exp\left[
i\frac{m}{4\hbar \tau}\left(1+2\tau\Omega\coth\Omega t\right)
\left(x^{\prime}\right)^2
\right]\\
\times\exp\left[-
i\left(
\frac{m\Omega}{\hbar\sinh\Omega t} \mathrm{e}^{t/2\tau}\ x^{\prime}+\frac{\Pi}{m}
\right)\left(x-\lambda\right)
\right].
\end{multline}
The Heisenberg picture position and
momentum operators and the propagator in the under-damping
regime can easily be found by doing the $\Omega\rightarrow i\Omega$.
 
\section{Two dimensional charged particle in time-dependent
 electric and magnetic fields}\label{twode}
In this section we show that the general method presented
in Sec. \ref{genmethod} can be extended
to obtain the evolution operator corresponding to the Hamiltonian
of a two-dimensional
charged particle ($-e$) confined to a quadratic potential subject
to an in-plane electric field and  perpendicular magnetic field. 
The Hamiltonian of such a system is given by
\begin{multline}
\hat H=\frac{1}{2m}\left({\hat p}_x+eA_x\right)^2+\frac{1}{2m}\left({\hat p}_y+eA_y\right)^2
-e\phi\\
+\frac{1}{2}K\left(\hat x^2+\hat y^2\right).\label{ham:magnetic}
\end{multline}
where $\hat x$, $\hat y$, $\hat p_x$ and $\hat p_y$ are the standard space and momentum operators in the $x-y$ plane.
The electron's charge is given by $e$  and
the scalar  and vector potentials are expressed in the
completely symmetric gauge  by
\begin{eqnarray}
\phi &=& -E_x\left(t\right)\hat x-E_y\left(t\right)\hat y,\\
A_x&=& -\frac{1}{2}B\left(t\right)\hat y,\\
A_y &=& \frac{1}{2}B\left(t\right)\hat x .
\end{eqnarray}
The mass $m$, the magnetic field $B$ and the coefficient $K$ may be
time-dependent.
Substituting the scalar and vector potentials in the
expression for the Hamiltonian and expanding,
we obtain \cite{PhysRevA.66.024103}
\begin{multline}
\hat H=\frac{1}{2m}\left({\hat p}_x^2+{\hat p}_y^2\right)
+\frac{1}{2}\left(K+\frac{e^2B^2}{4m}\right)\left(\hat x^2+\hat y^2\right)\\
+ \frac{eB}{2m}\left(\hat x\hat p_y-\hat y\hat p_x\right)
+eE_x\hat x+eE_y\hat y .\label{ham:mag}
\end{multline}
The structure shown by this Hamiltonian
(\ref{ham:mag}) suggests that the
set of generators that yields the corresponding closed Lie algebra
should be at least composed of the identity operator,
the generators listed in the previous section
($\hat \lambda_2$ to $\hat \lambda_6$)
for the $x$ and $y$ parts of (\ref{ham:mag}) and the angular momentum
$\hat L_z=\hat x\hat p_y-\hat y\hat p_x$.
However three more generators are needed in order to close
the algebra:
$\hat x\hat p_y+\hat y\hat p_x$, $\hat x\hat y$ and $\hat p_x\hat p_y$.
Thereby, the complete set is given
by $\hat \lambda_1=\hat 1$,
$\hat \lambda_2=\hat x$, $\hat \lambda_3=\hat p_x$,
$\hat \lambda_4=\hat x^2$, $\hat \lambda_5=\hat p_x^2$,
$\hat \lambda_6=\hat x\hat p_x+\hat p_x\hat x$,
$\hat \lambda_7=\hat y$, $\hat \lambda_8=\hat p_y$,
$\hat \lambda_9=\hat y^2$, $\hat \lambda_{10}=\hat p_y^2$,
$\hat \lambda_{11}=\hat y\hat p_y+\hat p_y\hat y$,
$\hat \lambda_{12}=\hat L_z=\hat x\hat p_y-\hat y\hat p_x$,
$\hat \lambda_{13}=\hat x\hat p_y+\hat y\hat p_x$,
$\hat \lambda_{14}=\hat x\hat y$
and
$\hat \lambda_{15}=\hat p_x\hat p_y$.
The algebra exhibited by this set of operators
is shown in Appendix \ref{gens:charge}.

As in the previous examples, we first deal with the linear terms through
the two-dimensional
generalization of the transformation shown in Eq. (\ref{gen:u1})
\begin{equation}
\hat U_1=\hat U_{1t}\hat U_{1x}\hat U_{1y},\label{twode:u1}
\end{equation}
where the $t$, $x$ and $y$ parts are given by
\begin{eqnarray}
\hat U_{1t} &=& \exp\left[\frac{i}{\hbar}S\left(t\right)\right],
\label{twode:u1t}\\
\hat U_{1x} &=&\exp\left[\frac{i}{\hbar}\Pi_x\left(t\right)\hat x\right]
   \exp\left[\frac{i}{\hbar}\lambda_x\left(t\right)\hat p_x\right],\\
\label{twode:u1x}
\hat U_{1y} &=&\exp\left[\frac{i}{\hbar}\Pi_y\left(t\right)\hat y\right]
   \exp\left[\frac{i}{\hbar}\lambda_y\left(t\right)\hat p_y\right].
\label{twode:u1y}
\end{eqnarray}
The corresponding transformation rules are
\begin{eqnarray}
\hat U_1\hat p_t\hat U_1^{\dagger} &=& \hat p_t+\dot S-\dot\lambda_x\Pi_x-\dot\lambda_y\Pi_y\nonumber \\
&&+\dot \Pi_x\hat x+\dot \lambda_x\hat p_x+\dot \Pi_y\hat y+\dot \lambda_y\hat p_y,\\
\hat U_1\hat x\hat U_1^{\dagger} &=& \hat x+\lambda_x,\\
\hat U_1\hat y\hat U_1^{\dagger} &=& \hat y+\lambda_y,\\
\hat U_1\hat p_x\hat U_1^{\dagger} &=& \hat p_x-\Pi_x,\\
\hat U_1\hat p_y\hat U_1^{\dagger} &=& \hat p_y-\Pi_y.
\end{eqnarray}

Under this transformation
the Floquet operator takes the form
\begin{multline}
\hat U_1\left(\hat H-\hat p_t\right)\hat U_1^{\dagger}
=\frac{1}{2m}\left(\hat p_x^2+\hat p_y^2\right)\\
+\frac{1}{2}\left(K+\frac{e^2B^2}{4m}\right)\left(\hat x^2+\hat y^2\right)
+\frac{eB}{2m}\left(\hat x\hat p_y-\hat y\hat p_x\right)\\
-\hat x\left[\frac{d}{dt}\frac{\partial  L}{\partial \dot \lambda_x}
-\frac{\partial L}{\partial\lambda_x}\right]
-\hat y\left[\frac{d}{dt}\frac{\partial  L}{\partial \dot \lambda_y}
-\frac{\partial  L}{\partial\lambda_y}\right]\\
+\hat p_x\left[\frac{d}{dt}\frac{\partial L}{\partial \dot \Pi_x}
-\frac{\partial  L}{\partial\Pi_x}\right]
+\hat p_y\left[\frac{d}{dt}\frac{\partial L}{\partial \dot \Pi_y}
-\frac{\partial  L}{\partial\Pi_y}\right]\\
-\hat p_t+ L-\dot S,\label{twode:flou1}
\end{multline}
yielding linear terms proportional to the Euler equations arising
from the classical Lagrangian
\begin{multline}
 L=\frac{1}{2m}\left(\Pi_x^2+\Pi_y^2\right)\\
+\frac{1}{2}\left(K+\frac{e^2B^2}{4m}\right)\left(\lambda_x^2+\lambda_ y^2\right)
+\frac{eB}{2m}\left(\lambda_y\Pi_x-\lambda_x\Pi_y\right)\\
+\dot\lambda_x \Pi_x+\dot\lambda_y \Pi_y
+eE_x\lambda_x+eE_y\lambda_y.
\end{multline}
In order to eliminate the linear terms in $\hat x$, $\hat y$, $\hat p_x$
and $\hat p_y$
we demand that the Euler equations vanish
\begin{eqnarray}
\frac{d}{dt}\frac{\partial  L}{\partial \dot \lambda_x}
-\frac{\partial L}{\partial\lambda_x} &=& -\left(K+\frac{e^2B^2}{4m}\right)\lambda_x
\label{charge:lambdax}
\nonumber\\
&&+\frac{eB}{2m}\Pi_y+\dot \Pi_x-eE_x=0 ,\\
\frac{d}{dt}\frac{\partial L}{\partial \dot \lambda_y}
- \frac{\partial L}{\partial\lambda_y} &=& -\left(K+\frac{e^2B^2}{4m}\right)\lambda_y
\nonumber\\
&&-\frac{eB}{2m}\Pi_x+\dot \Pi_y-eE_y=0 ,\\
\frac{d}{dt}\frac{\partial L}{\partial \dot \Pi_x}
-\frac{\partial L}{\partial\Pi_x}  &=& -\dot\lambda_x-\frac{\Pi_x}{m}-\frac{eB}{2m}\lambda_y=0 , 
\\
\frac{d}{dt}\frac{\partial L}{\partial \dot \Pi_y}
-\frac{\partial  L}{\partial\Pi_y} &=&  -\dot\lambda_y-\frac{\Pi_y}{m}+\frac{eB}{2m}\lambda_x=0 ,
\label{charge:Piy}
\end{eqnarray}
and $\dot S=L$. The Floquet operator
becomes
\begin{multline}
\hat U_1\left(\hat H-\hat p_t\right)\hat U_1^{\dagger}=\frac{1}{2m}\left(\hat p_x^2+\hat p_y^2\right)\\
+\frac{1}{2}\left(K+\frac{e^2B^2}{4m}\right)\left(\hat x^2+\hat y^2\right)\\
+\frac{eB}{2m}\left(\hat x\hat p_y-\hat y\hat p_x\right)-\hat p_t.
\end{multline}
The third term is proportional to the  $z$ projection of the angular momentum
$\hat L_z=\hat x\hat p_y-\hat y\hat p_x$
which is the generator
of rotations around the $z$ axis. We also notice that, besides the angular momentum
$L_z$, the first two terms given by the kinetic and potential
energy are also invariant under rotations. Hence the
next transformation is a rotation of the form
\begin{equation}
\hat U_2=\exp\left[\frac{i}{\hbar}\theta\left(t\right) \hat L_z\right]
=\exp\left[\frac{i}{\hbar}\theta\left(t\right)
\left(\hat x \hat p_y-\hat y\hat p_x\right) \right],
\label{twode:u2}
\end{equation}
with transformation rules given by
\begin{eqnarray}
\hat U_2\hat p_t\hat U_2^{\dagger} &=& \hat p_t+\dot \theta \left(\hat x \hat p_y-\hat y\hat p_x\right),\\
\hat U_2\hat x\hat U_2^{\dagger} &=& \cos\theta\hat x- \sin\theta\hat y,\\
\hat U_2\hat y\hat U_2^{\dagger} &=& \sin\theta\hat x+\cos\theta\hat y,\\
\hat U_2\hat p_x\hat U_2^{\dagger} &=& \cos\theta\hat p_x-\sin\theta \hat p_y,\\
\hat U_2\hat p_y\hat U_2^{\dagger} &=& \sin\theta\hat p_x +\cos\theta\hat p_y.
\end{eqnarray}
Under $\hat U_2$  the Floquet operator is transformed into
\begin{multline}
\hat U_2\hat U_1\left(\hat H-\hat p_t\right)\hat U_1^{\dagger}\hat U_2^{\dagger}
=\frac{1}{2m}\left(\hat p_x^2+\hat p_y^2\right)\\
+\frac{1}{2}\left(K+\frac{e^2B^2}{4m}\right)\left(\hat x^2+\hat y^2\right)\\
+\left(\frac{eB}{2m}-\dot\theta\right)\left(\hat x\hat p_y-\hat y\hat p_x\right)-\hat p_t.
\label{floquet:twode01}
\end{multline}
Here it is important to stress that all the elements in the previous
Floquet operator are invariant under rotations and therefore, the
only extra element introduced by the transformation arises from the
energy operator transformation rule.
By setting the restriction 
\begin{equation}
\dot \theta=eB/2m,\label{twode:u2theta}
\end{equation}
on the angle of rotation, the Floquet operator is transformed
into the one of two uncoupled harmonic oscillators
\begin{multline}
\hat U_2\hat U_1\left(\hat H-\hat p_t\right)\hat U_1^{\dagger}\hat U_2^{\dagger}
=\frac{1}{2m}\hat p_x^2+\frac{1}{2}\left(K+\frac{e^2B^2}{4m}\right)\hat x^2\\
+\frac{1}{2m}\hat p_y^2+\frac{1}{2}\left(K+\frac{e^2B^2}{4m}\right)\hat y^2-\hat p_t,
\label{charge:trans:u2}
\end{multline}
with time-dependent parameters $a=1/m$, $b=0$ and $c=K+e^2B^2/4m$.

It is clear how to proceed further: By repeating the procedure for the
general quadratic Hamiltonian for the $x$ and $y$ harmonic oscillators.
Even though in principle the two sets of transformations presented
in Secs. \ref{firstpath} and \ref{secondpath} are equivalent,
one is more effective than the other depending on the symmetries
of the system.
In  cases, such as the charged particle subject to
time-dependent magnetic field
with varying mass, the set of transformations  of Sec. \ref{firstpath}
yields closed and simple expressions for the transformation parameters
whereas the transformations of Sec. \ref{secondpath} give very complex ones.
However, some other systems, as the charged particle with variable mass
subject to constant magnetic field, 
are reduced using a smaller number of simple parameters
by means of the transformations of Sec. \ref{secondpath}.
In the latter case, solving Riccati differential equation
is conveniently avoided through the Arnold transformation.
These two cases are presented as
examples at the end of this section.

Let us now reduce the Floquet operator (\ref{floquet:twode01})
through the set of transformations from the first path (Sec. \ref{firstpath}).
In this case, the evolution operator is given by
\begin{equation}
\hat U^{\dagger}=\hat U_1^{\dagger}\hat U_2^{\dagger} \hat U_3^{\dagger} \hat U_4^{\dagger}
\hat U_5^{\dagger}\hat U_6^{\dagger},
\end{equation}
where $\hat U_1$ and $\hat U_2$ are given by Eqs. (\ref{twode:u1})
and (\ref{twode:u2}) respectively.
The remaining transformations are generalizations of
Eqs. (\ref{gen:u2}), (\ref{firstpath:u1}), (\ref{firstpath:u4}) and (\ref{firstpath:u5})
\begin{eqnarray}
\hat U_3 &=& \exp\left[\frac{i}{2\hbar}\gamma\left(\hat x\hat p_x+\hat p_x\hat x\right)
\right]\label{magnetic:fp:u3}\nonumber \\
&&\times\exp\left[\frac{i}{2\hbar}\gamma\left(\hat y\hat p_y+\hat p_y\hat y\right)\right],\\
\hat U_4 &=& \exp\left[\frac{i}{2\hbar}\alpha\Delta \hat x^2\right]
\exp\left[\frac{i}{2\hbar}\alpha\Delta \hat y^2\right],\\
\hat U_5 &=& \exp\left[\frac{i}{2\hbar}\phi\left(\hat x\hat p_x+\hat p_x\hat x\right)
\right]\nonumber \\
&&\times\exp\left[\frac{i}{2\hbar}\phi\left(\hat y\hat p_y+\hat p_y\hat y\right)\right],\\
\hat U_6 &=&
\exp\left[
\frac{i}{2\hbar }\beta\frac{\hat p_x^2}{\Delta}
\right]
\exp\left[
\frac{i}{2\hbar }\beta\frac{\hat p_y^2}{\Delta}
\right].\label{magnetic:fp:u6}
\end{eqnarray}
Since the $x$ and $y$ parts of the Floquet operator are symmetric,
the $x$ and $y$ part of these transformations have the
same parameters and they may be obtained from
the ordinary differential equations (\ref{gamma:path1}), (\ref{fp:eq1}),
(\ref{first:phi:u4}) and (\ref{first:beta:u5}). 

If instead we follow the procedure from Sec. \ref{secondpath}
the evolution operator is expressed as the product
\begin{equation}
\hat U^{\dagger}=\hat U_1^{\dagger}\hat U_2^{\dagger} \hat U_3^{\dagger} \hat U_4^{\dagger}
\hat U_5^{\dagger}\hat U_6^{\dagger}\hat U_7^{\dagger},\label{twode:evop}
\end{equation}
where, even though $\hat U_1$ and $\hat U_2$ are newly given
by (\ref{twode:u1})-(\ref{twode:u1y}) and (\ref{twode:u2}),
the remaining operators correspond to
the generalizations of  the transformations from the second path
(\ref{gen:u2}),
(\ref{gen:u3}), (\ref{gen:u4}), (\ref{gen:u5}) and (\ref{gen:u6})
\begin{eqnarray}
\hat U_3 &=& \exp\left[\frac{i}{2\hbar}\gamma\left(\hat x\hat p_x+\hat p_x\hat x\right)
\right]\nonumber\\
&&\times\exp\left[\frac{i}{2\hbar}\gamma\left(\hat y\hat p_y+\hat p_y\hat y\right)\right],
\\
\hat U_4 &=& \exp\left[\frac{i}{2\hbar}
\phi\left(\Delta\hat x^2+\frac{1}{\Delta}\hat p_x^2\right)\right]\nonumber\\
&&\times\exp\left[\frac{i}{2\hbar}
\phi\left(\Delta\hat y^2+\frac{1}{\Delta}\hat p_y^2\right)\right],\\
\hat U_5 &=& \exp\left[\frac{i}{2\hbar}\alpha\Delta \hat x^2\right]
\exp\left[\frac{i}{2\hbar}\alpha\Delta \hat y^2\right],
\\
\hat U_6 &=& \exp\left[\frac{i}{2\hbar}\varphi\left(\hat x\hat p_x+\hat p_x\hat x\right)
\right]\nonumber\\
&&\times\exp\left[\frac{i}{2\hbar}\varphi\left(\hat y\hat p_y+\hat p_y\hat y\right)\right],
\\
\hat U_7 &=& \exp\left[\frac{i}{2\hbar}\beta\frac{\hat p_x^2}{ \Delta}\right]
\exp\left[\frac{i}{2\hbar}\beta\frac{\hat p_y^2}{\Delta}\right].
\end{eqnarray}
The corresponding parameters may be obtained from
the ordinary differential equations (\ref{gamma:rest}), (\ref{u3:eq5}), (\ref{u4:eq5}),
(\ref{u5:tr7}) and (\ref{u6:tr5}).

Having derived the explicit form of the
evolution operator (\ref{twode:evop}) we obtain the Heisenberg
picture position and momentum operators as
\begin{equation}
\left[\begin{array}{l}
\hat x_{H}\left(t\right) \\
\hat y_{H}\left(t\right)\\
\hat p_{xH}\left(t\right)\\ 
\hat p_{yH}\left(t\right)\\ 
\end{array}\right]=
\mathbf{M}\left[\begin{array}{l}
\hat x \\
\hat y\\
\hat p_{x}\\
\hat  p_{y}
\end{array}\right]+
\left[\begin{array}{r}
\lambda_{x}\\
\lambda_{y}\\
-\Pi_{x}\\
-\Pi_{y}
\end{array}\right].\label{h:magnetic}
\end{equation}
In this case, $\mathbf{M}$ is a $4 \times 4$ matrix that has 
the following form
\begin{equation}
\mathbf{M}=\left[
\begin{array}{ll}
G_{qq}\mathbf{R} & G_{qp}\mathbf{R}\\
G_{pq}\mathbf{R} & G_{pp}\mathbf{R}
\end{array}
\right],
\end{equation}
with
\begin{equation}
\mathbf{R}=\left[\begin{array}{ll}
\cos\theta & -\sin\theta\\
\sin\theta & \cos\theta
\end{array}\right]
\end{equation}
the rotation matrix coming from the second transformation. 
Notice that, as in previous examples, $\mathbf{M}$ and $\mathbf{R}$ have a 
symplectic form.
The parameters $\lambda_x$, $\lambda_y$, $\Pi_x$ and $\Pi_y$ are
calculated from the classical differential equations of motion
(\ref{charge:lambdax})-(\ref{charge:Piy}), the rotation angle $\theta$
is given by (\ref{twode:u2theta}) and the coefficients
$G_{qq}$, $G_{qp}$, $G_{pq}$ and $G_{pp}$ can be obtained
by substituting the transformation parameters
in (\ref{coe:heis1})-(\ref{coe:heis4}) for the transformations
in Sec. \ref{firstpath} or (\ref{gqq})-(\ref{gpp}) for the transformations in
Sec. \ref{secondpath}.

For the first path, the Green function is calculated by placing the transformations
(\ref{twode:u1}), (\ref{twode:u2}) and (\ref{magnetic:fp:u3})-(\ref{magnetic:fp:u6})
in Eq. (\ref{prop}) and using the explicit form of the
propagators presented in Appendix
\ref{unitarytransformations}
\begin{multline}
G\left(x,y,t;x^\prime,y^\prime,0\right)=
\int dx_1dy_1dx_2dy_2\dots dx_5dy_5\\
\left\langle x,y\left\vert \hat U^{\dagger}_1\right\vert x_1,y_1\right\rangle
\left\langle x_1,y_1\left\vert \hat U^{\dagger}_2\right\vert x_2,y_2\right\rangle\\
\times \left\langle x_2,y_2\left\vert \hat U^{\dagger}_3\right\vert x_3,y_3\right\rangle
\dots 
\left\langle x_5,y_5\left\vert \hat U^{\dagger}_6\right\vert x^\prime,y^\prime\right\rangle\\
=\frac{\Delta}{2\pi\hbar\beta}\mathrm{e}^{-\phi-\gamma}
\exp\left(-i\frac{S}{\hbar}\right)\\
\times\exp\left[
-i\frac{\Pi_x}{\hbar}\left(x-\lambda_x\right)-i\frac{\Pi_y}{\hbar}\left(y-\lambda_y\right)
\right]\\
\times\exp\left\{i\frac{\Delta\mathrm{e}^{-2\gamma}}{2\hbar}
\left(\frac{\mathrm{e}^{-2\phi}}{\beta}-\alpha\right)
\left[\left(x-\lambda_x\right)^2+\left(y-\lambda_y\right)^2\right]
\right\}\\
\times\exp\left\{
i\frac{\Delta}{2\hbar\beta}\left[\left(x^\prime\right)^2+\left(y^\prime\right)^2\right]
\right\}\\
\times\exp\left\{-i\frac{\Delta\mathrm{e}^{-\phi-\gamma}}{\hbar\beta}
\left[\left(x-\lambda_x\right)\left(x^\prime\cos\theta-y^\prime\sin\theta\right)\right.\right.\\
\left. \frac{ }{ }\left.+\left(y-\lambda_y\right)\left(x^\prime\sin\theta+y^\prime\cos\theta\right)
\right]
\right\}.\label{prop:chpfp}
\end{multline}

Similarly, the second path's propagator
is calculated by gathering the explicit form of the seven
transformation propagators given in the Appendix \ref{unitarytransformations}
and performing the integral in
Eq. (\ref{prop}). This procedure yields
\begin{multline}
G \left(x,y,t; x^\prime,y^\prime,0 \right) =\int dx_1dy_1dx_2dy_2\dots dx_6dy_6\\
\left\langle x,y\left\vert \hat U^{\dagger}_1\right\vert x_1,y_1\right\rangle
\left\langle x_1,y_1\left\vert \hat U^{\dagger}_2\right\vert x_2,y_2\right\rangle\\
\times \left\langle x_2,y_2\left\vert \hat U^{\dagger}_3\right\vert x_3,y_3\right\rangle
\dots 
\left\langle x_6,y_6\left\vert \hat U^{\dagger}_7\right\vert x^\prime,y^\prime\right\rangle\\
=\frac{\Delta \exp\left(-\varphi-\gamma-i S/\hbar\right)}{4i\pi\hbar^2\beta\sin\phi}
\exp\left\{iw\left[\left(x^\prime\right)^2+\left(y^\prime\right)^2\right]\right\}\\
\times\exp\left\{iu\left[\left(x-\lambda_x\right)^2+\left(y-\lambda_y\right)^2\right]\right\}\\
\times \exp\left[
iq\left(x^\prime \cos\theta-y^\prime\sin\theta-\frac{\Pi_x}{q\hbar}\right)\left(x-\lambda_x\right)
\right.\\
\left.+iq\left(x^\prime \sin\theta+y^\prime\cos\theta-\frac{\Pi_y}{q\hbar}\right)\left(y-\lambda_y\right)
\right],
\end{multline}
where the functions $w$, $u$, $q$ and $l$
are given by Eqs. (\ref{w:function})-(\ref{l:function}).

\subsection{Charged particle in a time-varying magnetic field}\label{bt}
Here we consider a charged particle subject to a magnetic field of the
form $B=B_0\sin\omega t$.
Since there are no linear terms $\lambda_x=\lambda_y=\Pi_x=\Pi_y=S=0$.
By introducing this form of the magnetic field in Hamiltonian
(\ref{ham:magnetic}) and applying the first two transformations (\ref{twode:u1}) and
(\ref{twode:u2}) we are left with the Hamiltonian of two uncoupled
harmonic oscillators
\begin{multline}
\hat U_2\hat U_1\left(\hat H-\hat p_t\right)\hat U_1^{\dagger}\hat U_2^{\dagger}
=\frac{1}{2m}\hat p_x^2+\frac{1}{2}\frac{e^2B_0^2}{4m}\sin^2\omega t\ \hat x^2\\
+\frac{1}{2m}\hat p_y^2+\frac{1}{2}\frac{e^2B_0^2}{4m}\sin^2\omega t\ \hat  y^2-\hat p_t,
\end{multline}
where we identify $a=1/m$, $b=0$ and $c=e^2B_0^4\sin^2\omega t/4m=
e^2B_0^4\left(1-\cos2\omega t\right)/8m$. Given that in this
example the mass does not depend on time and consequently $a=$cnt. then $\gamma=0$.
Placing  $a$, $b$ and $c$ in the ordinary differential equations
(\ref{fp:eq1}), (\ref{phi:int}), (\ref{beta:int}) and (\ref{twode:u2theta})
we obtain the following solutions
\begin{eqnarray}
\alpha &=& -\frac{\omega C^\prime \left(\frac{\omega_c^2}{8\omega^2},
\frac{\omega_c^2}{16\omega^2},\omega t\right)}
{C\left(\frac{\omega_c^2}{8\omega^2},\frac{\omega_c^2}{16\omega^2},\omega t\right)}
=-\frac{\omega C^\prime\left(\omega t\right)}{C\left(\omega t\right)},\\
\phi &=& 
\ln\left[C\left(\frac{\omega_c^2}{8\omega^2},\frac{\omega_c^2}{16\omega^2},\omega t\right)\right]
=\ln\left[C\left(\omega t\right)\right],\\
\beta &=& \int_0^t\frac{ds}
{C^2\left(\frac{\omega_c^2}{8\omega^2},\frac{\omega_c^2}{16\omega^2},\omega s\right)}
=T\left(\omega t\right),\\
\theta &=& \frac{\omega_c}{\omega}\sin^2\frac{\omega t}{2}.
\end{eqnarray}
where $\omega_c = e B_0/m$ and $C$ and $C^\prime$ are the cosine Mathieu function and its
derivative which comply with $C\left(0\right)=1$ and $C^\prime\left(0\right)=0$.

The Heisenberg picture position and momentum operators are obtained by
replacing the previous parameters in (\ref{coe:heis1})-(\ref{coe:heis4})
 and (\ref{h:magnetic})
\begin{eqnarray}
\hat x_H  &=& C\left(\omega t\right)\left[\hat x \cos\left(\frac{\omega_c}{\omega}\sin^2\frac{\omega 
t}{2}\right)
-\hat y \sin\left(\frac{\omega_c}{\omega}\sin^2\frac{\omega t}{2}\right)\right]\nonumber\\
&&+\frac{T\left(\omega t\right)}{m}C\left(\omega t\right)
\left[\hat p_x \cos\left(\frac{\omega_c}{\omega}\sin^2\frac{\omega t}{2}\right)\right .\nonumber \\
&&\left.-\hat p_y \sin\left(\frac{\omega_c}{\omega}\sin^2\frac{\omega t}{2}\right)\right],\\
\hat y_H &=& C\left(\omega t\right)\left[\hat x \sin\left(\frac{\omega_c}{\omega}\sin^2\frac{\omega 
t}{2}\right)
+\hat y \cos\left(\frac{\omega_c}{\omega}\sin^2\frac{\omega t}{2}\right)\right]\nonumber\\
&&+\frac{T\left(\omega t\right)}{m}C\left(\omega t\right)
\left[\hat p_x \sin\left(\frac{\omega_c}{\omega}\sin^2\frac{\omega t}{2}\right)\right .\nonumber \\
&&\left.+\hat p_y \cos\left(\frac{\omega_c}{\omega}\sin^2\frac{\omega t}{2}\right)\right],\\
\hat p_{xH} &=& \left[\frac{1}{C\left(\omega t\right)}
+\omega C^\prime\left(\omega t\right)T\left(\omega t\right)\right]\left[\hat p_x 
\cos\left(\frac{\omega_c}{\omega}\sin^2\frac{\omega t}{2}\right)\right .\nonumber \\
&&\left.-\hat p_y \sin\left(\frac{\omega_c}{\omega}\sin^2\frac{\omega t}{2}\right)\right]
+m\omega C^\prime\left(\omega t\right)\left[\hat x \cos\left(\frac{\omega_c}{\omega}\sin^2\frac{\omega 
t}{2}\right)
\right.\nonumber \\
&&\left.-\hat y \sin\left(\frac{\omega_c}{\omega}\sin^2\frac{\omega t}{2}\right)\right],\\
\hat p_{yH} &=& \left[\frac{1}{C\left(\omega t\right)}
+\omega C^\prime\left(\omega t\right)T\left(\omega t\right)\right]\left[\hat p_x 
\sin\left(\frac{\omega_c}{\omega}\sin^2\frac{\omega t}{2}\right)\right .\nonumber \\
&&\left.+\hat p_y \cos\left(\frac{\omega_c}{\omega}\sin^2\frac{\omega t}{2}\right)\right]
+m\omega C^\prime\left(\omega t\right)\left[\hat x \sin\left(\frac{\omega_c}{\omega}\sin^2\frac{\omega 
t}{2}\right)
\right.\nonumber \\
&&\left.+\hat y \cos\left(\frac{\omega_c}{\omega}\sin^2\frac{\omega t}{2}\right)\right].
\end{eqnarray}

The propagator is calculated by introducing the explicit forms
of the transformation parameters in (\ref{prop:chpfp}) giving
\begin{multline}
G\left(x,y,t;x^\prime,y^\prime,0\right)=\frac{m}{2\pi\hbar T\left(\omega t\right)C\left(\omega t\right)}\\
\times\exp\left[i\frac{m}{2\pi\hbar}
\left(\frac{1}{T\left(\omega t\right)C^2\left(\omega t\right)}
+\frac{\omega C^\prime\left(\omega t\right)}{C\left(\omega t\right)}\right)\left(x^2+y^2\right)
\right]\\
\times\exp\left\{i\frac{m}{2\hbar T\left(\omega t\right)}\left[
\left(x^\prime\right)^2+\left(y^\prime\right)^2
\right]\right\}\\
\times\exp\left\{
-i\frac{m}{\hbar T\left(\omega t\right) C\left(\omega t\right)}\left[
\left(xx^\prime+yy^\prime\right)\cos\left(\frac{\omega_c}{\omega}\sin^2\frac{\omega 
t}{2}\right)\right.\right.\\
\left.\left.+\left(yx^\prime-xy^\prime\right)
\sin\left(\frac{\omega_c}{\omega}\sin^2\frac{\omega t}{2}\right)
\right]
\right\}.
\end{multline}

\subsection{Charged particle  in
time-dependent electric fields}\label{et}
The example treated in this section may be
of use in modelling single electron quantum dots  \cite{PhysRevA.81.052331},
or magneto transport in semiconductors under the influence of
an incident radiation  \cite{Inarrea201410}. In the latter application, the degree
of circular polarization plays an important role that may be elucidated through
the model presented in this section.
In order to allow the possibility of studying the effects of polarized incident light
we introduce the following form of the electric field
\begin{eqnarray}
E_x &=& E_{0x}+E_{1x}\sin\left(\omega t\right),\\
E_y &=& E_{0y}+E_{1y}\sin\left(\omega t+\zeta\right),
\end{eqnarray}
in Hamiltonian (\ref{ham:mag}) where $E_{0x}$ and $E_0y$ may be considered
bias electric fields and $\zeta$ controls the degree of circular
polarization of the incident radiation with electric field components
$E_{1x}$ and $E_{1y}$.
The first transformation to perform is (\ref{twode:u1}) that yields the
classical differential equations of motion (\ref{charge:lambdax})-(\ref{charge:Piy}).
The solution to these equations is obtained after a lengthy calculation
\begin{multline}
\lambda_x =-\frac{4 e E_{0x}}{\Gamma _-^2 m}
 +\frac{16 e E_{1y} \omega  \omega_c \cos \zeta \cos (t \omega )}{\Gamma ^4 m}\\
+\sin (t \omega ) \left(\frac{16 e E_{1x} \omega ^2}{\Gamma ^4 m}-\frac{4 \Gamma _-^2 e
   E_{1x}}{\Gamma ^4 m}-\frac{16 e E_{1y} \omega  \omega_c \sin \zeta}{\Gamma ^4
   m}\right)\\
+\sin \left(\frac{t \Omega }{2}\right) \sin \left(\frac{t \omega_c}{2}\right)
   \left(\frac{4 e E_{0x}
   \omega_c}{\Gamma _-^2 m \Omega }
   +\frac{32 e E_{1y} \omega ^3 \cos \zeta}{\Gamma ^4 m \Omega}\right.\\
   \left.
   -\frac{8 e E_{1y} \omega  \omega_c^2 \cos \zeta}{\Gamma ^4 m \Omega }-\frac{8 e E_{1y}
   \omega  \Omega  \cos \zeta}{\Gamma ^4 m}\right)\\
   +\cos \left(\frac{t \Omega }{2}\right) \cos \left(\frac{t\omega_c}{2}\right)
    \left(\frac{4 e E_{0x}}{\Gamma _-^2 m}-\frac{16 e E_{1y} \omega 
   \omega_c \cos \zeta}{\Gamma ^4 m}\right)\\
   +\cos \left(\frac{t \Omega}{2}\right) \sin \left(\frac{t \omega_c}{2}\right)
   \left(-\frac{4 e E_{0y}}{\Gamma _-^2m}
   -\frac{16 e E_{1x} \omega  \omega_c}{\Gamma ^4 m}\right.\\
   \left.+\frac{16 e E_{1y} \omega ^2 \sin \zeta}{\Gamma ^4 m}
   -\frac{4 \Gamma _-^2 e E_{1y} \sin \zeta}{\Gamma ^4 m}\right)\\
   +\sin \left(\frac{t \Omega }{2}\right) \cos \left(\frac{t \omega_c}{2}\right)
   \left(\frac{4 e E_{0y} \omega_c}{\Gamma _-^2 m \Omega }
   -\frac{32 e E_{1x} \omega ^3}{\Gamma ^4 m \Omega }\right. \\
  +\frac{8 \Gamma _+^2 eE_{1x} \omega }{\Gamma ^4 m \Omega }
   +\frac{16 e E_{1y} \omega ^2 \omega_c \sin \zeta}{\Gamma ^4 m \Omega }\\
    \left.+\frac{4 \Gamma _-^2 e E_{1y} \omega_c \sin \zeta}{\Gamma ^4 m \Omega}\right),
\end{multline}
\begin{multline}
\lambda_y=-\frac{4 e E_{0y}}{\Gamma_-^2 m}
+\cos (t \omega ) \left(-\frac{16 e E_{1x} \omega  \omega_c}{\Gamma ^4 m}\right.\\
\left.+\frac{16 eE_{1y} \omega ^2 \sin \zeta}{\Gamma ^4 m}
-\frac{4 \Gamma _-^2 e E_{1y} \sin \zeta}{\Gamma ^4 m}\right)\\
+\sin (t \omega ) \left(\frac{16 e E_{1y} \omega ^2 \cos \zeta}{\Gamma ^4 m}
-\frac{4 \Gamma_-^2 e E_{1y} \cos \zeta}{\Gamma ^4 m}\right)\\
+\sin \left(\frac{t \Omega }{2}\right) \cos \left(\frac{t \omega_c}{2}\right)
\left(-\frac{4 eE_{0x} \omega_c}{\Gamma _-^2 m \Omega }
-\frac{32 e E_{1y} \omega ^3 \cos \zeta}{\Gamma ^4m \Omega }\right.\\
\left.
+\frac{8 e E_{1y} \omega  \omega_c^2 \cos \zeta}{\Gamma ^4 m \Omega }
+\frac{8 eE_{1y} \omega  \Omega  \cos \zeta}{\Gamma ^4 m}\right)\\
+\cos \left(\frac{t \Omega }{2}\right) \sin\left(\frac{t \omega_c}{2}\right)
 \left(\frac{4 e E_{0x}}{\Gamma _-^2 m}
 -\frac{16 e E_{1y}\omega  \omega_c\cos \zeta}{\Gamma ^4 m}\right)\\
 +\cos \left(\frac{t \Omega }{2}\right) \cos\left(\frac{t \omega_c}{2}\right) 
 \left(\frac{4 e E_{0y}}{\Gamma _-^2 m}
 +\frac{16 e E_{1x}\omega  \omega_c}{\Gamma ^4 m}\right.\\
 \left.  -\frac{16 e E_{1y} \omega ^2 \sin \zeta}{\Gamma ^4 m}
   +\frac{4 \Gamma _-^2 e E_{1y} \sin \zeta}{\Gamma ^4 m}\right)\\
   +\sin \left(\frac{t \Omega }{2}\right) \sin\left(\frac{t \omega_c}{2}\right)
    \left(\frac{4 e E_{0y} \omega_c}{\Gamma _-^2 m \Omega}
    -\frac{32 e E_{1x} \omega ^3}{\Gamma ^4 m \Omega }
 \right. \\
     +\frac{8 \Gamma _+^2 e E_{1x} \omega }{\Gamma ^4 m   \Omega }
   +\frac{16 e E_{1y} \omega ^2 \omega_c \sin \zeta}{\Gamma ^4 m \Omega }\\
   \left.
   +\frac{4 \Gamma_-^2 e E_{1y} \omega_c \sin \zeta}{\Gamma ^4 m \Omega }\right),
\end{multline}
\begin{multline}
\Pi_x=\frac{2 eE_{0y}\omega_c}{\Gamma _-^2}\\
+\sin (t \omega )\left(\frac{8 eE_{1y}\omega ^2 \omega_c \cos \zeta}{\Gamma ^4}
+\frac{2 \Gamma _-^2 e E_{1y}\omega_c \cos \zeta}{\Gamma ^4}\right)\\
+\cos (t \omega ) \left(-\frac{16 eE_{1x}\omega ^3}{\Gamma ^4}
+\frac{4 \Gamma _+^2 eE_{1x}\omega }{\Gamma ^4}\right.\\
\left.+\frac{8 eE_{1y}\omega ^2 \omega_c \sin \zeta}{\Gamma^4}
+\frac{2 \Gamma _-^2 eE_{1y}\omega_c \sin \zeta}{\Gamma ^4}\right)\\
+\cos \left(\frac{t \Omega }{2}\right) \sin \left(\frac{t \omega_c}{2}\right) 
\left(-\frac{2 eE_{0x}\omega_c}{\Gamma _-^2}
-\frac{16 eE_{1y}\omega ^3 \cos \zeta}{\Gamma ^4}\right. \\
\left. +\frac{4eE_{1y}\omega  \Omega ^2 \cos \zeta}{\Gamma ^4}
+\frac{4 eE_{1y}\omega  \omega_c^2\cos \zeta}{\Gamma ^4}\right)\\
+\sin \left(\frac{t \Omega }{2}\right) \cos \left(\frac{t \omega_c}{2}\right) 
\left(\frac{2 eE_{0x}\Omega }{\Gamma _-^2}
-\frac{8 eE_{1y}\omega  \Omega \omega_c \cos \zeta}{\Gamma ^4}\right)\\
+\sin\left(\frac{t \Omega }{2}\right) \sin \left(\frac{t \omega_c}{2}\right)
 \left(-\frac{2 e E_{0y}\Omega }{\Gamma _-^2}
 -\frac{8 eE_{1x}\omega  \Omega  \omega_c}{\Gamma ^4}
 \right. \\
 \left.+\frac{8 e E_{1y}\omega ^2 \Omega  \sin \zeta}{\Gamma ^4}
 -\frac{2 \Gamma _-^2 eE_{1y}\Omega  \sin \zeta}{\Gamma^4}\right)\\
+\cos \left(\frac{t \Omega }{2}\right) \cos \left(\frac{t \omega_c}{2}\right)
\left(-\frac{2 eE_{0y}\omega_c}{\Gamma _-^2}
+\frac{16 eE_{1x}\omega ^3}{\Gamma^4}\right. \\
 \left.
-\frac{4 \Gamma _+^2 eE_{1x}\omega }{\Gamma ^4}
-\frac{8 eE_{1y}\omega ^2 \omega_c\sin \zeta}{\Gamma ^4}\right. \\
 \left.
-\frac{2 \Gamma _-^2 eE_{1y}\omega_c \sin \zeta}{\Gamma^4}\right).
\end{multline}
\begin{multline}
\Pi_y=
-\frac{2 e E_{0x} \omega_c}{\Gamma _-^2}\\
 +\sin (t \omega ) 
 \left(-\frac{8 e E_{1x} \omega ^2 \omega_c}{\Gamma ^4}
 -\frac{2\Gamma _-^2 e E_{1x} \omega_c}{\Gamma ^4}\right.\\ 
 \left.
 +\frac{16 e E_{1y} \omega ^3 \sin \zeta}{\Gamma^4}
 -\frac{4 e E_{1y} \omega  \Omega ^2 \sin \zeta}{\Gamma ^4}
 -\frac{4 e E_{1y} \omega \omega_c^2 \sin \zeta}{\Gamma ^4}\right)\\
 +\cos (t \omega )
 \left(\frac{4 \Gamma _+^2 e E_{1y}\omega  \cos \zeta}{\Gamma ^4}
 -\frac{16 e E_{1y} \omega ^3 \cos \zeta}{\Gamma ^4}\right)\\
+\cos \left(\frac{t \Omega }{2}\right) \cos \left(\frac{t \omega_c}{2}\right)
 \left(\frac{2 e E_{0x} \omega_c}{\Gamma _-^2}
 +\frac{16 e E_{1y}\omega ^3 \cos \zeta}{\Gamma ^4}\right.\\ 
 \left.
 -\frac{4 e E_{1y} \omega  \Omega ^2 \cos \zeta}{\Gamma ^4}
 -\frac{4e E_{1y} \omega  \omega_c^2 \cos \zeta}{\Gamma ^4}\right)\\
 +\sin \left(\frac{t \Omega }{2}\right) \sin \left(\frac{t \omega_c}{2}\right)
  \left(\frac{2 e E_{0x} \Omega }{\Gamma_-^2}
  -\frac{8 e E_{1y} \omega  \Omega  \omega_c \cos \zeta}{\Gamma ^4}\right)\\
  +\sin \left(\frac{t \Omega }{2}\right) \cos \left(\frac{t \omega_c}{2}\right)
   \left(\frac{2 e E_{0y}\Omega }{\Gamma _-^2}
   +\frac{8 e E_{1x} \omega  \Omega  \omega_c}{\Gamma ^4}\right.\\ 
 \left.
   -\frac{8 e E_{1y}\omega ^2 \Omega  \sin \zeta}{\Gamma ^4}
   +\frac{2 \Gamma _-^2 e E_{1y} \Omega  \sin \zeta}{\Gamma^4}\right)\\
   +\cos \left(\frac{t \Omega }{2}\right) \sin \left(\frac{t \omega_c}{2}\right)
   \left(-\frac{2 e E_{0y} \omega_c}{\Gamma _-^2}
   +\frac{16 e E_{1x} \omega ^3}{\Gamma^4}\right.\\ 
 \left.
   -\frac{4 \Gamma _+^2 e E_{1x} \omega }{\Gamma ^4}
   -\frac{8 e E_{1y} \omega ^2 \omega_c\sin \zeta}{\Gamma ^4}\right.\\ 
 \left.
   -\frac{2 \Gamma _-^2 e E_{1y} \omega_c \sin \zeta}{\Gamma^4}\right).
\end{multline}
where $\omega_c=eB/m$ is the cyclotron frequency, $\Omega^2=\left(4K+m\omega^2\right)/m$,
$\Gamma^4=16\omega^4+\left(\Omega^2-\omega_c^2\right)^2
-8\omega^2\left(\Omega+\omega_c\right)^2$,
$\Gamma_+^2=\Omega^2+\omega_c^2$ and
$\Gamma_-^2=\Omega^2-\omega_c^2$. 

Integrating Eq. (\ref{twode:u2theta}) yields the rotation angle given by
\begin{equation}
\theta=\frac{eB}{2m}t=\frac{\omega_c}{2}t.
\end{equation}
As mentioned above, this equations transform the
Floquet operator into the form of two uncoupled harmonic oscillators
with parameters $a=1/m$, $b=0$, $c=K+m(\omega_c/2)^2$, $d=0$, $e=0$ and
$g=0$.

In this example we follow the sequence of transformations
from the second path
presented in Sec. \ref{secondpath} which rapidly reduce the Floquet operator
avoiding the solution Riccati equation
in contrast to the procedure presented in Sec. \ref{firstpath} that yields
a larger number of parameters.
The remaining parameters $\gamma$, $\phi$, $\alpha$, $\varphi$ and $\beta$
are obtained from Eqs.  (\ref{gamma:rest}),
 (\ref{u3:eq5}), (\ref{u4:eq5}), (\ref{u5:tr7}) and (\ref{u6:tr5}).
These equations yield
\begin{eqnarray}
\gamma &=& 0, \\
\Delta &=& \frac{m\Omega}{2},\\
\phi &=& \frac{\Omega}{2}\ t, \\
\alpha &=& \varphi=\beta=0.
\end{eqnarray}
By replacing the previous parameters  in (\ref{gqq})-(\ref{gpp})
 and (\ref{h:magnetic})
we work out the explicit form of the Heisenberg picture space and momentum
operators
\begin{eqnarray}
\hat x_H &=& \cos\left(\frac{\Omega}{2}\ t\right)\left[\hat x\cos\left(\frac{\omega_c}{2}\ t\right)-\hat 
y\sin\left(\frac{\omega_c}{2}\ t\right)\right]\nonumber\\
&&+\frac{2}{m\Omega}\sin\left(\frac{\Omega}{2}\ t\right)\left[\hat p_x\cos\left(\frac{\omega_c}{2}\ 
t\right)-\hat p_y\sin\left(\frac{\omega_c}{2}\ t\right)\right]\nonumber\\
&&+\lambda_x,\\
\hat y_H &=& \cos\left(\frac{\Omega}{2}\ t\right)\left[\hat x\sin\left(\frac{\omega_c}{2}\ t\right)+\hat 
y\cos\left(\frac{\omega_c}{2}\ t\right)\right]\nonumber\\
&&+\frac{2}{m\Omega}\sin\left(\frac{\Omega}{2}\ t\right)\left[\hat p_x\sin\left(\frac{\omega_c}{2}\ 
t\right)+\hat p_y\cos\left(\frac{\omega_c}{2}\ t\right)\right]\nonumber\\
&&+\lambda_y,\\
\hat p_x &=& \cos\left(\frac{\Omega}{2}\ t\right)\left[\hat p_x\cos\left(\frac{\omega_c}{2}\ t\right)-\hat 
p_y\sin\left(\frac{\omega_c}{2}\ t\right)\right]\nonumber\\
&&-\frac{m\Omega}{2}\sin\left(\frac{\Omega}{2}\ t\right)\left[\hat x\cos\left(\frac{\omega_c}{2}\ 
t\right)-\hat y\sin\left(\frac{\omega_c}{2}\ t\right)\right]\nonumber\\
&&-\Pi_x,\\
\hat p_y &=& \cos\left(\frac{\Omega}{2}\ t\right)\left[\hat p_x\sin\left(\frac{\omega_c}{2}\ t\right)+\hat 
p_y\cos\left(\frac{\omega_c}{2}\ t\right)\right]\nonumber\\
&&-\frac{m\Omega}{2}\sin\left(\frac{\Omega}{2}\ t\right)\left[\hat x\sin\left(\frac{\omega_c}{2}\ 
t\right)+\hat y\cos\left(\frac{\omega_c}{2}\ t\right)\right]\nonumber\\
&&-\Pi_y.
\end{eqnarray}

In the calculation of the propagator, we only consider
the transformations $\hat U_1$, $\hat U_2$ and $\hat U_4$
since they have non-vanishing parameters
$\lambda_x$, $\lambda_y$, $\Pi_x$ and $\Pi_y$ and $\Delta$.
By inserting these transformations in (\ref{prop}),
the propagator takes the form 
\begin{multline}
G\left(x,y,t;x^\prime,y^\prime,0\right)=
\frac{m\Omega}{4\pi\hbar\sin\left(\frac{\Omega}{2}\ t\right) }
\mathrm{e}^{-iS/\hbar}\\
\times
\exp\left[-i\frac{\Pi_x}{\hbar}\left(x-\lambda_x\right)\right]
\exp\left[-i\frac{\Pi_y}{\hbar}\left(y-\lambda_y\right)\right]\\
\times\exp\left\{
i\frac{m\Omega}{4\hbar\sin\left(\frac{\Omega}{2}\ t\right) }\left[
\left(x^\prime\right)^2+\left(y^\prime\right)^2\right.\right.\\
\left.+\left(x-\lambda_x\right)^2
+\left(y-\lambda_y\right)^2
\right]\cos\left(\frac{\Omega}{2}\ t\right)\\
-2x^\prime\left(x-\lambda_x\right)\cos\left(\frac{\omega_c}{2}\ t\right)
-2x^\prime\left(y-\lambda_y\right)\sin\left(\frac{\omega_c}{2}\ t\right)\\
+2y^\prime\left(x-\lambda_x\right)\sin\left(\frac{\omega_c}{2}\ t\right)
-2y^\prime\left(y-\lambda_y\right)\cos\left(\frac{\omega_c}{2}\ t\right) \bigg\}.
\end{multline}

\section{Conclusions}\label{conclusions}
The Lie algebraic technics relay on the existence of
a set of generators that forms a closed algebra. If a given
Hamiltonian can be expressed as a linear combination
of these generators, the overall structure of its evolution
operator is known and takes the form of Eq. (\ref{unit0}).

We have applied the Lie algebraic approach to obtain the
evolution operator of the general harmonic oscillator
and the charged particle in time-dependent electric and magnetic
fields. The sets of operators that form closed Lie algebras
characterized by their structure constants
where established in each case.
Some particular examples of these two Hamiltonians were examined in detail.
The free particle in the presence of an external driving force was used
to introduce the Lie algebraic approach. Analytical expressions
for the evolution operator were provided for the potential of a radio
frequency ion trap as well as for  a forced harmonic oscillator
with varying mass.
The charged particle's evolution was studied under two sets of
different conditions. First we treated the case of a sinusoidally varying magnetic field
and second, we calculated explicit expressions for the evolution operator and propagator
of a particle in constant magnetic field
and time-dependent electric field.

The here obtained results may be used to tackle diverse problems
as squeezed states, radio frequency traps, and electronic transport
in two-dimensional lateral heterostructures under diverse conditions
of light excitation (linear polarization, circular polarization, etc.).
The methods developed so far can be extended to other types
of Hamiltonians, for example, a charged particle subject to time-dependent
electric and magnetic fields in a asymmetric confining parabolic potential.

We have observed that the Lie algebraic approach
is a powerful method that can be used to obtain the evolution operator
and propagator of a great variety of Hamiltonians. It reduces the
the difficulty of solving Schr\"odinger partial differential equation into
obtaining the solution of a system of coupled ordinary equations for the
transformation parameters.
Two possible shortcomings
of this method are that in general the solution 
of quite complex ordinary differential equations is needed,
and more important,
it requires a finite dimension set of operators that form a
closed algebra which in many cases is difficult to identify.

\acknowledgments
The authors would like to thank the ``Departamento de Ciencias B\'asicas UAM-A"
for the financial support and V. G. Ibarra-Sierra and J. C. Sandoval-Santana
would like to acknowledge the support
received from ``Becas de Posgrado UAM".

\appendix

\section{Generators of the Lie algebra}\label{liealgebra}
In this appendix we list the generators, and their corresponding
Lie algebras characterized by the structure constants for the
three examples treated in this paper: the linear potential,
the general quadratic Hamiltonian and the Hamiltonian of a charged particle
subject to electromagnetic fields.

\subsection{Generators for the linear potential Hamiltonian}\label{gens:lin}
The set of Hermitian operators that form the closed Lie algebra for
the Hamiltonian (\ref{ham:lin}) is given by
\begin{eqnarray}
\hat \lambda_1 &=& \hat 1,\\
\hat \lambda_2 &=& \hat x,\\
\hat \lambda_3 &=& \hat p,\\
\hat \lambda_4 &=& \hat p^2,
\end{eqnarray}
where $\hat 1$ is the identity operator.
The commutation relations arising from all the possible combinations
of the previous generators yield
\begin{eqnarray}
\left[\hat \lambda_1,\hat \lambda_2\right] &=& \left[\hat 1,\hat x\right] =0,\\
\left[\hat \lambda_1,\hat \lambda_3\right] &=& \left[\hat 1,\hat p\right] =0,\\
\left[\hat \lambda_1,\hat \lambda_4\right] &=& \left[\hat 1,\hat p^2\right] =0,\\
\left[\hat \lambda_2,\hat \lambda_3\right] &=& \left[\hat x,\hat p\right] =i\hbar \hat 1
=i\hbar \hat \lambda_1,\\
\left[\hat \lambda_2,\hat \lambda_4\right] &=& \left[\hat x,\hat p^2\right] =i\hbar 2\hat p
=i\hbar 2\hat \lambda_3,\\
\left[\hat \lambda_3,\hat \lambda_4\right] &=& \left[\hat p,\hat p^2\right] =0,
\end{eqnarray}
therefore the structure constants are $c_{2,3,1}=1$, $c_{2,4,3}=2$ all others begin zero
[see Eq. (\ref{strucons})].

\subsection{Generators for the general quadratic Hamiltonian}\label{gens:quadratic}
The structure of the quadratic Hamiltonian
(\ref{ham:quadratic}) suggests that
the closed algebra is given by
the set of operators
\begin{eqnarray}
\hat\lambda_1 &=&\hat 1,\label{lambda:1}\\
\hat\lambda_2 &=& \hat x,\label{lambda:x}\\
\hat\lambda_3 &=& \hat p,\label{lambda:p}\\
\hat\lambda_4 &=& \hat x^2,\label{lambda:x2}\\
\hat\lambda_5 &=& \hat p^2,\label{lambda:p2}\\
\hat\lambda_6 &=& \hat x\hat p+\hat p\hat x.\label{lambda:xp}
\end{eqnarray}
Indeed, the commutation relations for these operators yield
a closed algebra given by
\begin{eqnarray}
\left[\hat \lambda_1,\hat \lambda_2\right] &=& \left[\hat \lambda_1,\hat \lambda_3\right]
=\left[\hat \lambda_1,\hat \lambda_4\right] = \left[\hat \lambda_1,\hat \lambda_5\right]
\nonumber \\
&=&\left[\hat \lambda_1,\hat \lambda_6\right]=\left[\hat 1,\hat \lambda_i\right]=0,\\
\left[\hat \lambda_2,\hat \lambda_3\right] &=& \left[\hat x,\hat p\right]=i\hbar \hat 1
=i\hbar \hat \lambda_1,\label{commu:23}\\
\left[\hat \lambda_2,\hat \lambda_4\right] &=& \left[\hat x,\hat x^2\right]=0,\\
\left[\hat \lambda_2,\hat \lambda_5\right] &=& \left[\hat x,\hat p^2\right]=
i\hbar 2\hat p=i\hbar 2\lambda_3,\\
\left[\hat \lambda_2,\hat \lambda_6\right] &=& \left[\hat x,\hat x\hat p+\hat p\hat x\right]=
i\hbar 2\hat x=i\hbar 2\hat \lambda_2,\\
\left[\hat \lambda_3,\hat \lambda_4\right] &=& \left[\hat p,\hat x^2\right]=
-i\hbar 2\hat x=-i\hbar 2\hat\lambda_2,\\
\left[\hat \lambda_3,\hat \lambda_5\right] &=& \left[\hat p,\hat p^2\right]=0,\\
\left[\hat \lambda_3,\hat \lambda_6\right] &=& \left[\hat p,\hat x\hat p+\hat p\hat x\right]
=-i\hbar 2\hat p=-i\hbar 2\hat\lambda_3,\\
\left[\hat \lambda_4,\hat \lambda_5\right] &=& \left[\hat x^2,\hat p^2\right]=
i\hbar 2\left(\hat x\hat p+\hat p\hat x\right)=i\hbar 2\hat\lambda_6,\\
\left[\hat \lambda_4,\hat \lambda_6\right] &=& \left[\hat x^2,\hat x\hat p+\hat p\hat x\right]
=i\hbar 4\hat x^2=i\hbar 4\hat\lambda_4,\\
\left[\hat \lambda_5,\hat \lambda_6\right] &=& \left[\hat p^2,\hat x\hat p+\hat p\hat x\right]
=-i\hbar 4\hat p^2=-i\hbar 4\hat\lambda_5.\label{commu:56}
\end{eqnarray}
The structure constants are $c_{2,3,1}=1$, $c_{2,5,3}=2$, $c_{2,6,2}=2$,
$c_{3,4,2}=-2$, $c_{3,6,3}=-2$, $c_{4,5,6}=2$, $c_{4,6,4}=4$ and
$c_{5,6,5}=-4$, all others being zero.

\subsection{Generators for the Hamiltonian of a charged particle in electromagnetic
fields}\label{gens:charge}
The structure shown by the Hamiltonian
of a charged particle in electromagnetic fields
(\ref{ham:mag}) suggests that the
set of generators that yields the corresponding closed Lie algebra
should be composed of the identity operator,
the generators listed in the previous section ($\hat \lambda_2$ to $\hat \lambda_6$)
for the $x$ and $y$ parts of (\ref{ham:mag}) plus the angular momentum
$\hat L_z=\hat x\hat p_y-\hat y\hat p_x$. However, the algebra formed by this set
needs three more operators in order to be closed.
Thereby, the whole set of operators is given
by $\hat \lambda_1=\hat 1$,
$\hat \lambda_2=\hat x$, $\hat \lambda_3=\hat p_x$,
$\hat \lambda_4=\hat x^2$, $\hat \lambda_5=\hat p_x^2$,
$\hat \lambda_6=\hat x\hat p_x+\hat p_x\hat x$,
$\hat \lambda_7=\hat y$, $\hat \lambda_8=\hat p_y$,
$\hat \lambda_9=\hat y^2$, $\hat \lambda_{10}=\hat p_y^2$,
$\hat \lambda_{11}=\hat y\hat p_y+\hat p_y\hat y$,
$\hat \lambda_{12}=\hat L_z=\hat x\hat p_y-\hat y\hat p_x$,
$\hat \lambda_{13}=\hat x\hat p_y+\hat y\hat p_x$,
$\hat \lambda_{14}=\hat x\hat y$
and
$\hat \lambda_{15}=\hat p_x\hat p_y$.
Here is a summary of all the commutors arising from the generators
listed above, for convenient reference:
First, all the generators commute with the identity operator
\begin{multline}
\left[\hat \lambda_1,\hat \lambda_2\right]=\left[\hat \lambda_1,\hat \lambda_3\right]
=\left[\hat \lambda_1,\hat \lambda_4\right]=\left[\hat \lambda_1,\hat \lambda_5\right]
=\left[\hat \lambda_1,\hat \lambda_6\right]\\
=\left[\hat \lambda_1,\hat \lambda_7\right]
=\left[\hat \lambda_1,\hat \lambda_8\right]=\left[\hat \lambda_1,\hat \lambda_9\right]
=\left[\hat \lambda_1,\hat \lambda_{10}\right]\\
=\left[\hat \lambda_1,\hat \lambda_{11}\right]
=\left[\hat \lambda_1,\hat \lambda_{12}\right]
=\left[\hat \lambda_1,\hat \lambda_{13}\right]
=\left[\hat \lambda_1,\hat \lambda_{14}\right]\\
=\left[\hat \lambda_1,\hat \lambda_{15}\right]
=0.\label{gen:mat:first}
\end{multline}
Second, any generator from the $x$ part commutes with any generators
from the $y$ part
\begin{multline}
\left[\hat \lambda_2,\hat \lambda_{7}\right]=\left[\hat \lambda_2,\hat \lambda_{8}\right]
=\left[\hat \lambda_2,\hat \lambda_{9}\right]=\left[\hat \lambda_2,\hat \lambda_{10}\right]\\
=\left[\hat \lambda_2,\hat \lambda_{11}\right]
=\left[\hat \lambda_3,\hat \lambda_{7}\right]=\left[\hat \lambda_3,\hat \lambda_{8}\right]
=\left[\hat \lambda_3,\hat \lambda_{9}\right]\\
=\left[\hat \lambda_3,\hat \lambda_{10}\right]
=\left[\hat \lambda_3,\hat \lambda_{11}\right]
=\left[\hat \lambda_4,\hat \lambda_{7}\right]=\left[\hat \lambda_4,\hat \lambda_{8}\right]\\
=\left[\hat \lambda_4,\hat \lambda_{9}\right]=\left[\hat \lambda_4,\hat \lambda_{10}\right]
=\left[\hat \lambda_4,\hat \lambda_{11}\right]
=\left[\hat \lambda_5,\hat \lambda_{7}\right]\\
=\left[\hat \lambda_5,\hat \lambda_{8}\right]
=\left[\hat \lambda_5,\hat \lambda_{9}\right]=\left[\hat \lambda_5,\hat \lambda_{10}\right]
=\left[\hat \lambda_5,\hat \lambda_{11}\right]\\
=\left[\hat \lambda_6,\hat \lambda_{7}\right]=\left[\hat \lambda_6,\hat \lambda_{8}\right]
=\left[\hat \lambda_6,\hat \lambda_{9}\right]=\left[\hat \lambda_6,\hat \lambda_{10}\right]\\
=\left[\hat \lambda_6,\hat \lambda_{11}\right]=0.
\end{multline}
Third, the generators belonging to the $x$ coordinate must follow
the commutation rules (\ref{commu:23})-(\ref{commu:56}), therefore
\begin{eqnarray}
\left[\hat \lambda_2,\hat \lambda_3\right] &=& \left[\hat x,\hat p_x\right]=i\hbar \hat 1
=i\hbar \hat \lambda_1,\\
\left[\hat \lambda_2,\hat \lambda_4\right] &=& \left[\hat x,\hat x^2\right]=0,\\
\left[\hat \lambda_2,\hat \lambda_5\right] &=& \left[\hat x,\hat p_x^2\right]=
i\hbar 2\hat p_x=i\hbar 2\lambda_3,\\
\left[\hat \lambda_2,\hat \lambda_6\right] &=& \left[\hat x,\hat x\hat p_x+\hat p_x\hat x\right]=
i\hbar 2\hat x=i\hbar 2\hat \lambda_2,\\
\left[\hat \lambda_3,\hat \lambda_4\right] &=& \left[\hat p_x,\hat x^2\right]=
-i\hbar 2\hat x=-i\hbar 2\hat\lambda_2,\\
\left[\hat \lambda_3,\hat \lambda_5\right] &=& \left[\hat p_x,\hat p_x^2\right]=0,
\\
\left[\hat \lambda_3,\hat \lambda_6\right] &=& \left[\hat p_x,\hat x\hat p_x+\hat p_x\hat x\right]
=-i\hbar 2\hat p_x\nonumber\\
&=&-i\hbar 2\hat\lambda_3,\\
\left[\hat \lambda_4,\hat \lambda_5\right] &=& \left[\hat x^2,\hat p_x^2\right]=
i\hbar 2\left(\hat x\hat p_x+\hat p_x\hat x\right)\nonumber\\
&=&i\hbar 2\hat\lambda_6,\\
\left[\hat \lambda_4,\hat \lambda_6\right] &=& \left[\hat x^2,\hat x\hat p_x+\hat p_x\hat x\right]
=i\hbar 4\hat x^2=i\hbar 4\hat\lambda_4,\\
\left[\hat \lambda_5,\hat \lambda_6\right] &=& \left[\hat p_x^2,\hat x\hat p_x+\hat p_x\hat x\right]
=-i\hbar 4\hat p_x^2\nonumber\\
&=&-i\hbar 4\hat\lambda_5.
\end{eqnarray}
Similarly for the $y$ coordinate we have
\begin{eqnarray}
\left[\hat \lambda_7,\hat \lambda_8\right] &=& \left[\hat y,\hat p_y\right]=i\hbar \hat 1
=i\hbar \hat \lambda_1,\\
\left[\hat \lambda_7,\hat \lambda_9\right] &=& \left[\hat y,\hat y^2\right]=0,\\
\left[\hat \lambda_7,\hat \lambda_{10}\right] &=& \left[\hat y,\hat p_y^2\right]=
i\hbar 2\hat p_y=i\hbar 2\hat\lambda_8,\\
\left[\hat \lambda_7,\hat \lambda_{11}\right] &=& 
\left[\hat y,\hat y\hat p_y+\hat p_y\hat y\right]=
i\hbar 2\hat y=i\hbar 2\hat \lambda_7,\\
\left[\hat \lambda_8,\hat \lambda_9\right] &=& \left[\hat p_y,\hat y^2\right]=
-i\hbar 2\hat y=-i\hbar 2\hat\lambda_7,\\
\left[\hat \lambda_8,\hat \lambda_{10}\right] &=& \left[\hat p_y,\hat p_y^2\right]=0,
\end{eqnarray}
\begin{eqnarray}
\left[\hat \lambda_8,\hat \lambda_{11}\right] &=& \left[\hat p_y,\hat y\hat p_y+\hat p_y\hat y\right]
=-i\hbar 2\hat p_y\nonumber\\
&=&-i\hbar 2\hat\lambda_8,\\
\left[\hat \lambda_9,\hat \lambda_{10}\right] &=& \left[\hat y^2,\hat p_y^2\right]=
i\hbar 2\left(\hat y\hat p_y+\hat p_y\hat y\right)\nonumber\\
&=&i\hbar 2\hat\lambda_{11},\\
\left[\hat \lambda_9,\hat \lambda_{11}\right] &=& \left[\hat y^2,\hat y\hat p_y+\hat p_y\hat y\right]
=i\hbar 4\hat y^2=i\hbar 4\hat\lambda_9,\\
\left[\hat \lambda_{10},\hat \lambda_{11}\right] &=&
\left[\hat p_y^2,\hat y\hat p_y+\hat p_y\hat y\right]
=-i\hbar 4\hat p_y^2\nonumber\\
&=&-i\hbar 4\hat\lambda_{10}.
\end{eqnarray}
The remaining operators need to be calculated independently
%
\begin{eqnarray}
\left[\hat \lambda_2,\hat \lambda_{12}\right]&=&\left[\hat x,\hat L_z\right]
=-i\hbar \hat y=-i\hbar\hat\lambda_7,\\
\left[\hat \lambda_2,\hat \lambda_{13}\right]&=&\left[\hat x,\hat x\hat p_y+\hat y\hat p_x\right]
=i\hbar \hat y=i\hbar\hat\lambda_7,\\
\left[\hat \lambda_2,\hat \lambda_{14}\right]&=&\left[\hat x,\hat x\hat y\right]=0,\\
\left[\hat \lambda_2,\hat \lambda_{15}\right]&=&\left[\hat x,\hat p_x\hat p_y\right]=
i\hbar p_y=i\hbar \hat\lambda_8,
\end{eqnarray}
%
\begin{eqnarray}
\left[\hat \lambda_3,\hat \lambda_{12}\right]&=&\left[\hat p_x,\hat L_z\right]
=-i\hbar \hat p_y=-i\hbar\hat\lambda_8,\\
\left[\hat \lambda_3,\hat \lambda_{13}\right]&=&
\left[\hat p_x, \hat x\hat p_y+\hat y\hat p_x \right]
=-i\hbar \hat p_y=-i\hbar\hat\lambda_8,\\
\left[\hat \lambda_3,\hat \lambda_{14}\right]&=&\left[\hat p_x, \hat x\hat y \right]
=-i\hbar \hat y=-i\hbar\hat\lambda_7,\\
\left[\hat \lambda_3,\hat \lambda_{15}\right]&=&\left[\hat p_x, \hat p_x\hat p_y \right]
=0,
\end{eqnarray}
%
\begin{eqnarray}
\left[\hat \lambda_4,\hat \lambda_{12}\right]&=&\left[\hat x^2,\hat L_z\right]
=-i\hbar 2\hat x\hat y=-i\hbar 2\hat\lambda_{14},\\
\left[\hat \lambda_4,\hat \lambda_{13}\right]&=&\left[\hat x^2,\hat x\hat p_y+\hat y\hat p_x\right]
=i\hbar 2\hat x\hat y=i\hbar 2\hat\lambda_{14},\\
\left[\hat \lambda_4,\hat \lambda_{14}\right]&=&\left[\hat x^2,\hat x\hat y\right]
=0,\\
\left[\hat \lambda_4,\hat \lambda_{15}\right]&=&\left[\hat x^2,\hat p_x\hat p_y\right]
=i\hbar 2\hat x\hat p_y\nonumber \\
&=&i\hbar \left(\hat\lambda_{12}+\hat\lambda_{13}\right),
\end{eqnarray}
%
\begin{eqnarray}
\left[\hat \lambda_5,\hat \lambda_{12}\right]&=&\left[\hat p_x^2,\hat L_z\right]
=-i\hbar 2\hat p_x\hat p_y=-i\hbar 2\hat\lambda_{15},\\
\left[\hat \lambda_5,\hat \lambda_{13}\right]&=&\left[\hat p_x^2,\hat x\hat p_y+\hat y\hat p_x\right]
\nonumber \\
&=&-i\hbar 2\hat p_x\hat p_y=-i\hbar 2\hat\lambda_{15},\\
\left[\hat \lambda_5,\hat \lambda_{14}\right]&=&\left[\hat p_x^2,\hat x\hat y\right]
=-i\hbar 2\hat y\hat p_x\nonumber \\
&=&i\hbar \left(\hat\lambda_{12}-\hat\lambda_{13}\right),\\
\left[\hat \lambda_5,\hat \lambda_{15}\right]&=&\left[\hat p_x^2,\hat p_x\hat p_y\right]
=0,
\end{eqnarray}
%
\begin{eqnarray}
\left[\hat \lambda_6,\hat \lambda_{12}\right]&=&\left[\hat x\hat p_x+\hat p_x \hat x,\hat L_z\right]
\nonumber\\
&=&-i\hbar 2\left(\hat x\hat p_y+\hat y\hat p_x\right)
=-i\hbar 2\hat\lambda_{13},\\
\left[\hat \lambda_6,\hat \lambda_{13}\right]
&=&\left[\hat x\hat p_x+\hat p_x \hat x,\hat x\hat p_y+\hat y\hat p_x\right]
\nonumber\\
&=&-i\hbar 2 \hat L_z
=-i\hbar 2\hat\lambda_{12},\\
\left[\hat \lambda_6,\hat \lambda_{14}\right]
&=&\left[\hat x\hat p_x+\hat p_x \hat x,\hat x\hat y\right]
\nonumber\\
&=&-i\hbar 2 \hat x\hat y
=-i\hbar 2\hat\lambda_{14},\\
\left[\hat \lambda_6,\hat \lambda_{15}\right]
&=&\left[\hat x\hat p_x+\hat p_x \hat p_x,\hat p_x\hat p_y\right]
\nonumber\\
&=&i\hbar 2 \hat p_x\hat p_y
=i\hbar 2\hat\lambda_{15},
\end{eqnarray}
%
\begin{eqnarray}
\left[\hat \lambda_7,\hat \lambda_{12}\right]&=&\left[\hat y,\hat L_z\right]
=i\hbar \hat x=i\hbar\hat\lambda_2,\\
\left[\hat \lambda_7,\hat \lambda_{13}\right]&=&\left[\hat y,\hat x\hat p_y+\hat y\hat p_x\right]
=i\hbar \hat x=i\hbar\hat\lambda_2,\\
\left[\hat \lambda_7,\hat \lambda_{14}\right]&=&\left[\hat y,\hat x\hat y\right]=0,\\
\left[\hat \lambda_7,\hat \lambda_{15}\right]&=&\left[\hat y,\hat p_x\hat p_y\right]
=i\hbar \hat p_x=i\hbar \hat \lambda_3,
\end{eqnarray}
%
\begin{eqnarray}
\left[\hat \lambda_8,\hat \lambda_{12}\right]&=&\left[\hat p_y,\hat L_z\right]
=i\hbar \hat p_x=i\hbar\hat\lambda_3,\\
\left[\hat \lambda_8,\hat \lambda_{13}\right]&=&\left[\hat p_y,\hat x\hat p_y+\hat y\hat p_x\right]
=-i\hbar \hat p_x=-i\hbar\hat\lambda_3,\\
\left[\hat \lambda_8,\hat \lambda_{14}\right]&=&\left[\hat p_y,\hat x\hat y\right]
=-i\hbar \hat x=-i\hbar\hat\lambda_2,\\
\left[\hat \lambda_8,\hat \lambda_{15}\right]&=&\left[\hat p_y,\hat p_x\hat p_y\right]
=0,
\end{eqnarray}
%
\begin{eqnarray}
\left[\hat \lambda_9,\hat \lambda_{12}\right]&=&\left[\hat y^2,\hat L_z\right]
=i\hbar 2\hat x \hat y=i\hbar 2\hat\lambda_{14},\\
\left[\hat \lambda_9,\hat \lambda_{13}\right]&=&\left[\hat y^2,\hat x\hat p_y+\hat y\hat p_x\right]
=i\hbar 2\hat x\hat y=i\hbar 2\hat\lambda_{14},\\
\left[\hat \lambda_9,\hat \lambda_{14}\right]&=&\left[\hat y^2,\hat x\hat y\right]=0,\\
\left[\hat \lambda_9,\hat \lambda_{15}\right]&=&\left[\hat y^2,\hat p_x \hat p_y\right]
=i\hbar \hat y \hat p_x\nonumber \\
&=& i\hbar \left(\hat\lambda_{13}-\hat\lambda_{12}\right),
\end{eqnarray}
%
\begin{eqnarray}
\left[\hat \lambda_{10},\hat \lambda_{12}\right]&=&\left[\hat p_y^2,\hat L_z\right]
=i\hbar 2\hat p_x \hat p_y=i\hbar 2\hat\lambda_{15},\\
\left[\hat \lambda_{10},\hat \lambda_{13}\right]&=&\left[\hat p_y^2,\hat x\hat p_y+\hat y\hat p_x\right]
\nonumber \\
&=&-i\hbar 2\hat p_x\hat p_y=-i\hbar 2\hat\lambda_{15},\\
\left[\hat \lambda_{10},\hat \lambda_{14}\right]&=&\left[\hat p_y^2,\hat x\hat y\right]
\nonumber \\
&=&-i\hbar 2\hat x \hat p_y=-i\hbar \left(\hat\lambda_{12}+\hat\lambda_{13}\right),\\
\left[\hat \lambda_{10},\hat \lambda_{15}\right]&=&\left[\hat p_y^2,\hat p_x\hat p_y\right]
=0,
\end{eqnarray}
%
\begin{eqnarray}
\left[\hat \lambda_{11},\hat \lambda_{12}\right]&=&
\left[\hat y\hat p_y+\hat p_y \hat y,\hat L_z\right]
\nonumber\\
&=&i\hbar 2\left(\hat x\hat p_y+\hat y\hat p_x\right)
=i\hbar 2\hat\lambda_{13},\\
\left[\hat \lambda_{11},\hat \lambda_{13}\right]
&=&\left[\hat y\hat p_y+\hat p_y \hat y,\hat x\hat p_y+\hat y\hat p_x\right]
\nonumber \\
&=& i\hbar 2 \hat L_z=i\hbar 2\hat \lambda_{12},\\
\left[\hat \lambda_{11},\hat \lambda_{14}\right]
&=&\left[\hat y\hat p_y+\hat p_y \hat y,\hat x\hat y\right]
\nonumber \\
&=& -i\hbar 2 \hat x\hat y=-i\hbar \hat 2\lambda_{14},\\
\left[\hat \lambda_{11},\hat \lambda_{15}\right]
&=&\left[\hat y\hat p_y+\hat p_y \hat y,\hat p_x\hat p_y\right]
\nonumber \\
&=& i\hbar 2 \hat p_x\hat p_y=i\hbar \hat 2\lambda_{15},
\end{eqnarray}
%
\begin{eqnarray}
\left[\hat \lambda_{12},\hat \lambda_{13}\right]&=&\left[\hat L_z,\hat x\hat p_y+\hat y \hat p_x\right]
\nonumber \\
&=& i\hbar \left(\hat y\hat p_y+\hat p_y\hat y-\hat x\hat p_x-\hat p_x\hat x\right)
\nonumber \\
&=& i\hbar \left(\hat\lambda_{11}-\hat\lambda_6\right),\\
\left[\hat \lambda_{12},\hat \lambda_{14}\right]&=&\left[\hat L_z,\hat x\hat y\right]
\nonumber \\
&=& i\hbar \left(\hat y^2-\hat x^2\right)
= i\hbar \left(\hat \lambda_9-\hat \lambda_4\right),\\
\left[\hat \lambda_{12},\hat \lambda_{15}\right]&=&\left[\hat L_z,\hat p_x\hat p_y\right]
\nonumber \\
&=& i\hbar \left(\hat p_y^2-\hat p_x^2\right)
= i\hbar \left(\hat \lambda_{10}-\hat \lambda_5\right),
\end{eqnarray}
%
\begin{eqnarray}
\left[\hat \lambda_{13},\hat \lambda_{14}\right]&=&\left[\hat x\hat p_y+\hat y \hat p_x,\hat x\hat y\right]
\nonumber \\
&=& -i\hbar \left(\hat x^2+\hat y^2\right)
= -i\hbar \left(\hat \lambda_4+\hat \lambda_9\right),\\
\left[\hat \lambda_{13},\hat \lambda_{15}\right]
&=&\left[\hat x\hat p_y+\hat y \hat p_x,\hat p_x\hat p_y\right]
\nonumber \\
&=& i\hbar \left(\hat p_x^2+\hat p_y^2\right)
= i\hbar \left(\hat \lambda_5+\hat \lambda_{10}\right),\\
\left[\hat \lambda_{14},\hat \lambda_{15}\right]
&=&\left[\hat x\hat y,\hat p_x\hat p_y\right]\nonumber \\
&=& i\hbar \frac{1}{2} \left(\hat x\hat p_x+\hat p_x\hat x+\hat y\hat p_y+\hat p_y\hat 7\right)
\nonumber \\
&=& i\hbar \frac{1}{2}\left(\hat \lambda_6+\hat \lambda_{11}\right).
\label{gen:mat:last}
\end{eqnarray}
The structure constants inferred from the $15!2!/\left(15-2\right)!2!=105$ commutors
in Eqs. (\ref{gen:mat:first})-(\ref{gen:mat:first})
are $c_{2,3,1}=1$, $c_{2,5,3}=2$, $c_{2,6,2}=2$, $c_{3,4,2}=-2$, $c_{3,6,3}=-2$,
$c_{4,5,6}=2$, $c_{4,6,4}=4$, $c_{5,6,5}=-4$, $c_{7,8,1}=1$, $c_{7,10,8}=2$,
$c_{7,11,7}=2$, $c_{8,9,7}=-2$, $c_{8,11,8}=-2$, $c_{9,10,11}=2$, $c_{9,11,9}=4$,
$c_{10,11,10}=-4$, $c_{2,12,7}=-1$, $c_{2,13,7}=1$, $c_{2,15,8}=1$,
$c_{3,12,8}=-1$, $c_{3,13,8}=-1$, $c_{3,14,7}=-1$, $c_{4,12,14}=-2$,
$c_{4,13,14}=2$, $c_{4,15,12}=1$, $c_{4,15,13}=1$, $c_{5,12,15}=-2$,
$c_{5,13,15}=-2$, $c_{5,14,12}=1$, $c_{5,14,13}=-1$, $c_{6,12,13}=-2$,
$c_{6,13,12}=-2$, $c_{6,14,14}=-2$, $c_{6,15,15}=2$,
$c_{7,12,2}=1$, $c_{7,13,2}=1$, $c_{7,15,3}=1$, $c_{8,12,3}=1$, $c_{8,13,3}=-1$,
$c_{8,14,2}=-1$, $c_{9,12,14}=2$, $c_{9,13,14}=2$, $c_{9,15,12}=-1$, $c_{9,15,13}=1$,
$c_{10,12,15}=2$, $c_{10,13,15}=-2$, $c_{10,14,12}=-1$, $c_{10,14,13}=-1$,
$c_{11,12,13}=2$, $c_{11,13,12}=2$, $c_{11,14,14}=-2$, $c_{11,15,15}=2$,
$c_{12,13,6}=-1$, $c_{12,13,11}=1$, $c_{12,14,4}=-1$, $c_{12,14,9}=1$,
$c_{12,15,5}=-1$, $c_{12,15,10}=1$, $c_{13,14,4}=-1$, $c_{13,14,9}=-1$,
$c_{13,15,5}=1$, $c_{13,15,10}=1$, $c_{14,15,6}=1/2$, $c_{14,15,11}=1/2$,
all others being zero.

\section{Unitary tranformations}\label{unitarytransformations}

The following  sections are devoted to presenting the
unitary transformations generated by the operators in the previous
appendices used along the paper to reduce the
different Floquet operators.
The general form of each transformation is accompanied
by the transformation rules, i. e. the explicit forms of
$\hat U \hat p_t \hat U^\dagger$, $\hat U \hat x \hat U^\dagger$ 
and $\hat U \hat p \hat U^\dagger$.
The transformation's Green functions
are also presented in the sections to follow.

Since it is widely used in the following appendices
we enunciate the next commutation relation.
If the commutor 
\begin{equation}
 \left[\hat{A},\hat{B}\right]=\hat{C},
\end{equation}
commutes with the operators $\hat A$ and $\hat B$, i. e.
\begin{equation}
\left[\hat{A},\hat{C}\right]=\left[\hat{B},\hat{C}\right]=0.
\end{equation}
then it follows that
\begin{equation}\label{relation-cuan}
\left[\hat{A},F(\hat{B})\right]=\left[\hat{A},\hat{B}\right]\frac{\partial
F(\hat{B})}{\partial\hat{B}},
\end{equation}
provided that $F$ is an analytical function.

%
%
\subsection{Unitary transformation generated by $\hat x$ and $\hat p$.
Shift in space and momentum}\label{unitarytranslation}

This transformation shifts the
space and momentum operators by time-dependent functions.
It is generated by $\hat \lambda_1$, $\hat \lambda_2$ and
$\hat \lambda_3$ in Eqs. (\ref{lambda:1}), (\ref{lambda:x}) and
(\ref{lambda:p}) as follows
\begin{multline}
\hat{U} =\hat{U}_t \hat{U}_x \hat{U}_p = \exp{\left[ \frac{i}{\hbar} S(t) \right]}\\
\times \exp{ \left[ \frac{i}{\hbar} \Pi(t) \hat{x}\right]} \exp{\left[\frac{i}{\hbar} 
\lambda (t) \hat{p} \right]} \ .
\end{multline}
The transformation rules for the space, momentum and energy operators 
can be worked out by inserting commutors
\begin{eqnarray}
\hat{U} \hat{x} \hat{U}^{\dagger}  &=& \hat{x} + 
\hat U_p\left[\hat{x},\hat{U}_p^{\dagger} \right],\\
\hat{U} \hat{p} \hat{U}^{\dagger} &=& \hat{p} + 
\hat{U}_x \left[\hat{p},\hat{U}^{\dagger}_x \right], \\
\hat{U} \hat{p}_t \hat{U}^{\dagger} &=& \hat{p}_t +\hat U_t \left[\hat{p}_t,\hat{U_t^{\dagger} }\right],
\end{eqnarray}
and using relation
(\ref{relation-cuan}) as follows
\begin{eqnarray}
\hat{U} \hat{x} \hat{U}^{\dagger} &=&
\hat{x} + \hat{U}_p\left[\hat{x},\hat{p}\right]
\frac{\partial \hat{U}^{\dagger}_p}{\partial p} = \hat{x} +  \lambda \ ,\\
\hat{U} \hat{p} \hat{U}^{\dagger} &=& 
\hat{p} +\hat U_x \left[\hat{p},\hat{x}\right] 
\frac{\partial}{\partial x} \hat{U}^{\dagger}_x = \hat{p} -  \Pi \ ,\\
\hat{U} \hat{p}_t \hat{U}^{\dagger} &=& \hat{p}_t + \hat{U}_t \left[\hat p_t,t\right]
\frac{\partial}{\partial t} \hat{U_t^{\dagger}} \nonumber\\
&=& \hat{p}_t + \dot{S} - \dot{\lambda} \Pi + \dot{\Pi} \hat{x} + \dot{\lambda} \hat{p} \ .
\end{eqnarray}

The propagator for this transformation is given by
\begin{multline}
\left\langle x \left\vert U^{\dagger} \right\vert x^{\prime} \right\rangle = \exp{\left[ -\frac{i S}{\hbar} \right]} \exp{ \left[-\frac{i \Pi}{\hbar} x^{\prime} \right]} \\
\times\delta \left( x - x^{\prime} - \lambda \right),
\end{multline}
where $\delta$ is the Dirac delta distribution.
%
%
\subsection{Transformation generated by $\hat x\hat p+\hat p\hat x$.
 Dilation}\label{unitarydilation}
Dilations are generated by $\hat\lambda_6$ in Eq. (\ref{lambda:xp}).
The explicit form of this transformation is given by
\begin{equation}
\hat{U} = \exp{ \left[ \frac{i}{2 \hbar} \gamma (t) \left( \hat{x} \hat{p} + \hat{p} \hat{x}\right) \right]}.
\end{equation}
In order to get the transformation rules for the position and momentum operators we define
\begin{eqnarray}
\hat X\left(\gamma\right) &=& \hat U\hat x\hat U^{\dagger},\\
\hat P\left(\gamma\right) &=& \hat U\hat p\hat U^{\dagger},
\end{eqnarray}
and compute the derivatives with respect to the transformation parameter $\gamma$
\begin{eqnarray}
\frac{\partial}{\partial \gamma} \hat{X}(\gamma)  &=& \frac{i}{2 \hbar}
\hat{U} \left[\hat{x} \hat{p}+\hat{p} \hat{x}, \hat{x} \right] \hat{U}^{\dagger} =  \hat{X} \ , \\
\frac{\partial}{\partial \gamma} \hat{P}(\gamma)  &=& \frac{i}{2 \hbar}
\hat{U} \left[\hat{x} \hat{p}+\hat{p} \hat{x}, \hat{p} \right] \hat{U}^{\dagger}= - \hat{P} \ .
\end{eqnarray}
The solution to this pair of differential equations together with
 initial conditions
$\tilde{X}(0)=\hat{x}$ and $\tilde{P}(0)=\hat{p}$ yields the standard transformation rules
for dilations
\begin{eqnarray}
\hat{X}= \hat{x} \ e^{\gamma}  , \\
\hat{P}= \hat{p} \ e^{-\gamma} .
\end{eqnarray}
The transformation rule for the energy operator is easily
calculated by inserting a commutor
\begin{equation}
\hat{U} \hat{p}_t \hat{U}^{\dagger}  = \hat{p}_t +  \hat U\left[\hat p_t,\hat U^\dagger\right],
\end{equation} 
and using relation (\ref{relation-cuan}) as follows
\begin{equation}
\hat{U} \hat{p}_t \hat{U}^{\dagger}  = \hat{p}_t + i \hbar \ \hat{U} \frac{\partial}{\partial t} \hat{U^{\dagger}}=  \hat{p_t} + \frac{\dot{\gamma}}{2}  \left( \hat{x} \hat{p} + \hat{p} \hat{x} \right).
\end{equation}
The corresponding propagator is given by
\begin{equation}
\left\langle x \left\vert U^{\dagger} \right\vert x^{\prime} \right\rangle = e^{-\frac{\gamma}{2}} \ \delta \left( e^{-\gamma} x - x^{\prime} \right) .
\end{equation}

\subsection{Transformation generated by $\hat x^2$}\label{unitaryx2}
Here we analyze the transformations generated by
$\hat \lambda_4$ in Eq. (\ref{lambda:x2}).
\begin{equation}
\hat{U} =   \exp{\left[{i\alpha(t) \frac{\Delta  \hat{x}^2}{2 \hbar}}\right]} \ .
\end{equation}
Since it depends explicitly on the position operator, the position operator
itself remain unaltered under its action
\begin{equation}
\hat{U} \hat{x} \hat{U^{\dagger}}= \hat{x} \ .
\end{equation}

The momentum and energy transformation rules are easily obtained
by inserting a commutor and using relation (\ref{relation-cuan}) as follows
\begin{eqnarray}
\hat{U} \hat{p} \hat{U^{\dagger}} &=& \hat{p}
+  \hat{U} [\hat{p},\hat{x}] \frac{\partial \hat{U^{\dagger}}}{\partial x} = \hat{p} - \alpha  \Delta \hat{x}  \ ,
\label{x2:rule}\\
\hat{U} \hat{p_t} \hat{U^{\dagger}} &=& \hat{p}_t + i \hbar \ \hat{U} \frac{\partial}{\partial t} \hat{U^{\dagger}}= \hat{p}_t + \frac{\dot{\alpha} \Delta}{2} \hat{x}^2 \ .
\end{eqnarray}
The propagator associated to this transformation is easily calculated by
using  (\ref{x2:rule})
\begin{equation}
\left\langle x \left\vert U^{\dagger} \right\vert x^{\prime} \right\rangle
= \exp{\left({-\frac{i \alpha \Delta}{2 \hbar} {x^{\prime}}^2} \right)} \ \delta \left( x - x^{\prime} \right).
\end{equation}

\subsection{Transformation generated by $\hat p^2$}\label{unitaryp2}

This transformation is generated by $\hat \lambda_5$ in Eq. (\ref{lambda:p2}) given by
\begin{equation}
\hat{U} = \exp{\left[i\beta (t) \frac{\hat{p}^2}{2 \hbar}   \right]}
\ .
\end{equation}
Under the action of $\hat U$, $\hat p$ remains unaltered
since these two operators trivially commute 
\begin{equation}
\hat{U} \hat{p} \hat{U^{\dagger}}= \hat{p} \ .
\end{equation}
The energy and position transformations rules  are easily worked out by
inserting a commutor and using the relation (\ref{relation-cuan}) as follows
\begin{eqnarray}
\hat{U} \hat{x} \hat{U^{\dagger}} &=& \hat{x} +\hat{U} [\hat{x},\hat{p}]
 \frac{\partial \hat{U^{\dagger}}}{\partial \hat{p}}= \hat{x} + \beta \hat{p}  \ , \\
\hat{U} \hat{p}_t \hat{U}^{\dagger} &=& \hat{p}_t + i \hbar \ \hat{U} \frac{\partial}{\partial t} 
\hat{U^{\dagger}}= \hat{p}_t + \frac{\dot{\beta}}{2} \hat{p}^2 \ .
\end{eqnarray}

This transformation's propagator is given by
\begin{equation}
\left\langle x \left\vert U^{\dagger} \right\vert x^{\prime} \right\rangle = \frac{1}{\sqrt{2 \pi \hbar \beta}} \exp{ \left[\frac{i}{2 \hbar \beta} (x - x^{\prime})^2 \right]} 
\end{equation}

\subsection{Transformation generated by $\Delta \hat x^2+\hat p^2/\Delta$.
Arnold transformation}\label{arnold:sec}

The Arnold transformation is generated
by a linear combination of $\hat\lambda_4$ and $\hat\lambda_5$
in Eqs. (\ref{lambda:x2}) and (\ref{lambda:p2}).
This transformation's explicit form is given by
\begin{equation}
{\hat U}\left(t\right)=\exp\left[\frac{i}{2\hbar}\phi\left(t\right)
\left(\Delta {\hat x}^2+\frac{1}{\Delta}{\hat p}^2\right)
\right],\label{arnold:u}
\end{equation}
where $\Delta$ is a constant
that yields a unit-less $\phi$ transformation parameter.

The transformation rule for the energy operator is readily
calculated by inserting a commutor as follows
\begin{multline}
\hat U \hat p_t \hat U^{\dagger}=\hat p_t+\hat U\left[\hat p_t,\hat U^{\dagger}\right]
=\hat p_t+\hat U\left[\hat p_t,t\right]\frac{\partial \hat U^{\dagger}}{\partial t}\\
 ={\hat p}_t+\frac{1}{2}\dot\phi\left(\frac{1}{\Delta}{\hat p}^2
  + \Delta {\hat x}^2\right).
\end{multline}
In order to obtain the transformation rules for the position and momentum operators we define
\begin{eqnarray}
\hat X\left(\phi\right) &=& \hat U\hat x\hat U^{\dagger},\\
\hat P\left(\phi\right) &=& \hat U\hat p\hat U^{\dagger},
\end{eqnarray}
and compute the derivatives with respect
to the transformation parameter
\begin{eqnarray}
\frac{d}{d\phi}\hat X\left(\phi\right) &=& \frac{i}{2\hbar}\hat U\left[\hat x, 
\Delta {\hat x}^2+\frac{1}{\Delta}{\hat p}^2
\right]\hat U^{\dagger}\nonumber \\
&=&-\frac{\hat P\left(\phi\right)}{\Delta},\\
\frac{d}{d\phi}\hat P\left(\phi\right) &=& \frac{i}{2\hbar} \hat U\left[\hat p,
\Delta {\hat x}^2+\frac{1}{\Delta}{\hat p}^2
\right]\hat U^{\dagger}\nonumber \\
&=&\Delta \hat X\left(\phi\right).
\end{eqnarray}
The solution to this system of differential equations together with the
boundary conditions $\hat X\left(0\right)=\hat x$ and $\hat P\left(0\right)=\hat p$
yields the transformation rules
\begin{eqnarray}
{\hat U}\hat x{\hat U}^\dagger &=& \hat x \cos\phi+\frac{1}{\Delta}\hat p\sin\phi,\\
{\hat U}\hat p{\hat U}^\dagger &=& \hat p \cos\phi-\Delta \hat x \sin \phi.
\end{eqnarray}
The Arnold's transformation propagator is given by
\begin{multline}
\left\langle x \left\vert U^{\dagger} \right\vert x^{\prime} \right\rangle
= \sqrt{\frac{\Delta}{2 \pi \hbar \sin{\phi}}}\\
\times\exp{\frac{i \Delta}{2 \hbar \sin{\phi}} \left[ ({x^{\prime}}^2 + x^2) \cos{\phi}-2 x x^{\prime}\right]}. 
\end{multline}
 
\subsection{Transformations generated by $\hat L_z$. Rotations}
Rotations are generated by $\hat \lambda_{12}$ in Sec. \ref{gens:charge}.
The transformation is given by
\begin{equation}
\hat U=\exp\left[i\frac{\theta\left(t\right)}{\hbar}\hat L_z\right].
\end{equation}
The transformation rule for the energy operator is easily
calculated by inserting a commutor and using (\ref{relation-cuan})
as follows
\begin{multline}
\hat U \hat p_t \hat U^{\dagger}=\hat p_t+\hat U\left[\hat p_t,\hat U^{\dagger}\right]
=\hat p_t+\hat U\left[\hat p_t,t\right]\frac{\partial \hat U^{\dagger}}{\partial t}\\
 ={\hat p}_t+\dot\theta\hat L_z.
\end{multline}
In order to obtain the transformation rules for the position and momentum operators
we define
\begin{eqnarray}
\hat X\left(\theta\right) &=& \hat U\hat x\hat U^{\dagger},\\
\hat Y\left(\theta\right) &=& \hat U\hat y\hat U^{\dagger},
\end{eqnarray}
and calculate their derivatives with respect
to the rotation angle $\theta$
\begin{eqnarray}
\frac{d}{d\theta}\hat X\left(\theta\right) &=& \frac{i}{\hbar}\hat U\left[\hat x, 
\hat L_z
\right]\hat U^{\dagger}=Y\left(\theta\right),\\
\frac{d}{d\theta}\hat Y\left(\theta\right) &=& \frac{i}{\hbar}\hat U\left[\hat y, 
\hat L_z
\right]\hat U^{\dagger}=-X\left(\theta\right).
\end{eqnarray}
The solution to this system of differential equations together with the
boundary conditions $\hat X\left(0\right)=\hat x$
and $\hat Y\left(0\right)=\hat y$
yields the transformation rules
\begin{eqnarray}
{\hat U}\hat x{\hat U}^\dagger &=&\cos\theta \hat x -\sin\theta\hat y ,\label{trlz:x}\\
{\hat U}\hat y{\hat U}^\dagger &=& \sin\theta\hat x +\cos\theta\hat y .\label{trlz:y}
\end{eqnarray}
Following a similar procedure for the momentum operators we obtain
the rules
\begin{eqnarray}
{\hat U}\hat p_x{\hat U}^\dagger &=&\cos\theta \hat p_x -\sin\theta\hat p_y ,\\
{\hat U}\hat p_y{\hat U}^\dagger &=& \sin\theta\hat p_x +\cos\theta\hat p_y.
\end{eqnarray}

The propagator can readily be obtained from
the transformation rules (\ref{trlz:x}) and (\ref{trlz:y})
giving
\begin{multline}
\left\langle x,y \left\vert \hat U^{\dagger}\right\vert x^\prime, y^\prime\right\rangle=
\delta\left(x-x^\prime \cos\theta+y^\prime \sin\theta\right)\\
\times\delta\left(y-x^\prime \sin\theta-y^\prime \cos\theta\right).
\end{multline}

%

\end{document}